\documentclass[useAMS,usenatbib]{mn2e}

\usepackage{graphicx}
\usepackage{times}

\def\kms{km ${\rm s}^{-1}$}

\def\ch2{$\chi^2$}

\def\kms {\hbox{${\rm km\ s}^{-1}$}}

\def\scm  {$\hbox{{\rm cm}}^{-2}$}    

\def\MOLH {\hbox{${\rm H}_2$}}  

\def\rr {\rightarrow}

\def \AL {$\alpha $}     
\def \HI {H{\sc \,i}}
\def \WpHz {W Hz$^{-1}$}
\def\lapp{\ifmmode\stackrel{<}{_{\sim}}\else$\stackrel{<}{_{\sim}}$\fi}
\def\gapp{\ifmmode\stackrel{>}{_{\sim}}\else$\stackrel{>}{_{\sim}}$\fi}

\title[Redshifted Radio Absorption Lines II]{A Survey for Redshifted Molecular and Atomic Absorption Lines\\{\Large II. Associated \HI, OH and millimetre lines in the {\boldmath $z\gapp3$} Parkes Quarter-Jansky Flat-spectrum Sample}}
  \author[S. J. Curran et al.]{S. J. Curran$^{1}$\thanks{E-mail: sjc@phys.unsw.edu.au}, M. T. Whiting$^{1,2}$, T. Wiklind$^{3,4}$, J. K. Webb$^{1}$, M. T. Murphy$^{5,7}$, C. R. Purcell$^{1,6}$\\
$^{1}$School of Physics, University of New South Wales, Sydney NSW 2052, Australia\\
$^{2}$CSIRO Australia Telescope National Facility, PO Box 76, Epping NSW 1710, Australia\\
$^{3}$Onsala Space Observatory, S-439 92 Onsala, Sweden\\
$^{4}$Space Telescope Science Institute, Baltimore, Maryland 21218, USA\\
$^{5}$Institute of Astronomy, Madingley Road, Cambridge CB3 0HA, UK\\
$^{6}$Jodrell Bank Centre for Astrophysics, University of Manchester, Alan
Turing Building, Oxford Road, Manchester M13 9PL, UK\\
$^{7}$Centre for Astrophysics and Supercomputing, Swinburne University
of Technology, PO Box 218, Hawthorn, VIC 3122, Australia}

\begin{document}

\date{Accepted ---. Received ---; in original form ---}

\pagerange{\pageref{firstpage}--\pageref{lastpage}} \pubyear{2008}

\maketitle

\label{firstpage}

\begin{abstract}

We present the results of a $z\geq2.9$ survey for \HI\ 21-cm and
molecular absorption in the hosts of radio quasars using the Giant
Metrewave Radio Telescope and the Tidbinbilla 70-m telescope. Although
the atomic gas has been searched to limits capable of detecting most
known absorption systems, no \HI\ was detected in any of the ten
sources. Previously published searches, which are overwhelmingly at
redshifts of $z\lapp1$, exhibit a 42\% detection rate (31 out of 73
sources), whereas the inclusion of our survey yields a 17\% detection
rate (2 out of 12 sources) at $z>2.5$. We therefore believe that our
high redshift selection is responsible for our exclusive
non-detections, and find that at ultra-violet luminosities of $L_{\rm
  UV}\gapp10^{23}$ \WpHz, 21-cm absorption has never been detected. We
also find this to not only apply to our targets, but also those at low
redshift exhibiting similar luminosities, giving zero detections out
of a total of 16 sources over $z=0.24$ to $3.8$. This is in contrast
to the $L_{\rm UV}\lapp10^{23}$ \WpHz\ sources where there is a near
50\% detection rate of 21-cm absorption.

The mix of 21-cm detections and non-detections is currently attributed
to orientation effects, where according to unified schemes of active
galactic nuclei, 21-cm absorption is more likely to occur in sources
designated as radio galaxies (type-2 objects, where the nucleus is
viewed through dense obscuring circumnuclear gas) than in quasars
(type-1 objects, where we have a direct view to the nucleus). However,
due to the exclusively high ultra-violet luminosities of our targets
it is not clear whether orientation effects {\em alone} can wholly
account for the distribution, although there exists the possibility
that the large luminosities are indicative of a changing demographic
of galaxy types. 
We also find that below
luminosities of $L_{\rm UV}\sim10^{23}$ \WpHz, {\em both} type-1 and
type-2 objects have a 50\% likelihood of exhibiting 21-cm absorption.

Finally, we do not detect molecular gas in any of the sources.
The lack of
\HI\ absorption, combined with the results from Paper~I, suggest these
sources are not conducive to high molecular abundances.

\end{abstract}

\begin{keywords}
radio lines: galaxies -- galaxies: active --  quasars: absorption lines
 -- cosmology: observations -- galaxies: abundances -- galaxies: high redshift
\end{keywords}

\section{Introduction}
\label{intro}

Redshifted molecular and atomic absorption lines can provide excellent
probes of the contents and nature of the early Universe. In
particular, with redshifted radio and microwave lines we can
investigate the gaseous content and large-scale structure as well any
possible variations in the values of the fundamental constants at
large look-back times. However, these are currently rare, with only 67
\HI\ 21-cm absorption systems at $z\gapp0.1$ known, comprising of 37
associated systems and 30 intervening (Table \ref{t1}). For molecular
absorption in the radio band the situation is considerably worse with
only five redshifted OH 18-cm systems currently known 
\citep{cdn99,kc02a,kcdn03}, four of which also exhibit a plethora of molecular
absorption lines in the millimetre regime (see \citealt{cw98b}).
\begin{table*}
\centering
\begin{minipage}{110mm}
\caption{The known redshifted ($z_{\rm abs}\gapp0.1$) \HI~21-cm
absorbers (updated from Paper I, \citealt{cwm+06}). Absorber types are:
BLRG--broad line radio galaxy, CSO--compact symmetric object,
CSS--compact steep spectrum source, DLA--damped Lyman-$\alpha$ absorption
system (Mg{\sc \,ii}--DLA candidate), GPS--gigahertz peaked spectrum
source, HFP -- high frequency peaker galaxy, Lens--gravitational lens,
OHM--OH megamaser, Red--red quasar, RG--radio galaxy. The number of
each type is given, as well as the absorption redshift and column
density ranges.
\label{t1}}
\begin{tabular}{@{}l c c c c @{}} 
\hline
Reference      & Type    & No.& $z_{\rm abs}$ & $N_{\rm HI}$ [\scm
                                                                  ]\\
\hline
\multicolumn{5}{c}{ASSOCIATED ABSORBERS}\\
\hline
\citet{mir89}  & RG      & 2  & 0.10 \& 0.12 & $2$ \& $6\times10^{18}.\,(T_{\rm s}/f)$ \\
\citet{vke+89} & RG      & 2  & 0.06 \& 0.10 & $11$ \& $8\times10^{18}.\,(T_{\rm s}/f)$ \\
\citet{ubc91}  & RG      & 1  & 3.40       & $3\times10^{18}.\,(T_{\rm s}/f)$ \\
\citet{cps92}  & Red     & 1  & 0.25       & $1\times10^{19}.\,(T_{\rm s}/f)$ \\
\citet{cmr+98} & Red     & 3  & 0.58--0.67 & $0.8 - 8\times10^{19}.\,(T_{\rm s}/f)$ \\
\citet{mcm98}  & Red     & 1  & 2.64 & $8\times10^{18}.\,(T_{\rm s}/f)$ \\
\citet{ptc99}  & CSO     & 1  & 0.10       & $3\times10^{19}.\,(T_{\rm s}/f)$\\
\citet{ptf+00} & CSO     & 1  & 0.25       & $5\times10^{18}.\,(T_{\rm s}/f)$ \\
\citet{ida03} & CSS/Red & 1  & 1.19       & $4\times10^{19}.\,(T_{\rm s}/f)$ \\
\citet{vpt+03} & BLRG    & 1  & 0.22       & $7\times10^{17}.\,(T_{\rm s}/f)$\\
...            & CSS     & 7  & 0.19--0.80 & $0.1-2\times10^{18}.\,(T_{\rm s}/f)$\\
...            & GPS     & 10 & 0.08--0.65 & $0.07-3\times10^{19}.\,(T_{\rm s}/f)$ \\
...            & RG      & 1  & 0.24       & $1\times10^{18}.\,(T_{\rm s}/f)$ \\
\citet{pbdk05} & OHM     & 1  & 0.22       & $6 \times10^{18}.\,(T_{\rm s}/f)$ \\
\citet{cwm+06} & RG      & 1  & 0.10       & $4\times10^{19}.\,(T_{\rm s}/f)$\\
\citet{gs06}   & RG      & 1  & 0.08       & $6\times10^{18}.\,(T_{\rm s}/f)$ \\
\citet{gss+06} & CSS     & 1  & 0.17       & $5\times10^{18}.\,(T_{\rm s}/f)$ \\
\citet{omd06}$^{*}$  & HFP     & 1  & 0.67       & $8\times10^{19}.\,(T_{\rm s}/f)$\\
\hline
\multicolumn{5}{c}{INTERVENING ABSORBERS}\\
\hline
\citet{cry93}            & Lens & 1  & 0.69 & $1\times10^{19}.\,(T_{\rm s}/f)$ \\
\citet{lrj+96}           & Lens & 1  & 0.19 & $\approx2\times10^{18}.\,(T_{\rm s}/f)$ \\

\citet{cdn99}            & Lens & 1  & 0.89 & $1\times10^{19}.\,(T_{\rm s}/f)$ \\
\citet{kb03}             & Lens & 1  & 0.76 & $1\times10^{19}.\,(T_{\rm s}/f)$ \\
\citet{kc02}$^{\dagger}$ & DLA  & 15 & 0.09--2.04 &  $0.02-6\times10^{19}.\,(T_{\rm s}/f)$ \\ 
\citet{dgh+04}           & DLA  & 1  & 0.78 & $2\times10^{19}.\,(T_{\rm s}/f)$ \\
\citet{kse+06}           & DLA  & 1  & 2.35 & $4\times10 ^{17}.\,(T_{\rm s}/f)$ \\ 
\citet{kcl06}            & DLA  & 1  & 3.39 &  $1\times10 ^{18}.\,(T_{\rm s}/f)$\\
\citet{gsp+06}           & Mg{\sc \,ii} & 3 & 1.17--1.37 & $0.4 - 2\times10^{18}.({T_{\rm s}}/{f})$\\
\citet{cdbw07}           &           Lens & 1 & 0.96 & $2\times10 ^{18}.\,(T_{\rm s}/f)$\\
\citet{ctm+07}           & DLA  & 1  & 0.66 & $4\times10 ^{18}.\,(T_{\rm s}/f)$ \\
\citet{ykep07}           & DLA  & 1  & 2.29 & $2\times10 ^{18}.\,(T_{\rm s}/f)$ \\
Zwaan et al. (in prep.) & Mg{\sc \,ii} & 2 & $\sim0.6$ & --\\ 
\hline
\end{tabular}
{Notes: $^{*}$Also J1407+2827, which is counted as one of the 10 GPSs of \citet{vpt+03}.
$^{\dagger}$See the paper for the full reference
list and \protect\citet{cmp+03} for the calculated column densities. 
Note that since PKS 1413+135 is an associated system, it has
been included in the top panel.}
\end{minipage}
\end{table*}

From our previous survey for radio absorption lines in the hosts of
the sources in the Parkes Half-Jansky Flat-spectrum Sample (PHFS), one
\HI\ absorption system was clearly detected (out of five searched) and
one OH system tentatively detected (of the 13 searched), both at
$z_{\rm abs}\sim0.1$ (Paper I). Upon examination of the previous
detections, we concluded:
\begin{itemize}
  \item For the \HI\ 21-cm absorption there is no overwhelming correlation
    between the line strength and the optical--near-infrared ($V-K$) colour.

    \item However, for the OH 18-cm absorption there is a clear
    relationship, thus suggesting that the reddening of these quasars
    is due to dust, the amount of which is correlated with the
    molecular abundance. Furthermore, for the molecular absorbers:
    \begin{itemize}
        \item All of the absorption lines were found at redshifts were
	  absorption (usually \HI) was already known to occur\footnote{Except in the
	case of PKS 1830--211 where a gravitational lens of undetermined
	redshift was previously known \citep{snps90}. The redshift was finally 
	determined through a 14 GHz wide
  spectral scan of the 3-mm band \citep{wc96}.}.
      
	  \item In all cases the absorption occurs towards flat
      spectrum radio sources, suggesting compact radio sources
      and thus a large effective coverage by the absorber.

    \end{itemize}
\end{itemize}
Unfortunately, prior to the analysis undertaken in Paper I, not all of
the above criteria were fully formulated, and so we have also
targetted sources for which only the last criterion is satisfied.  In
this paper we present the results of a survey for redshifted atomic
and molecular absorption within the hosts of sources selected from the
Parkes Quarter-Jansky Flat-spectrum Sample (PQFS,
\citealt{jws+02}). 

Due to the search for coincident molecular
absorption, we initially prioritised sources in which the HCO$^+$
$0\rr1$ transition would be redshifted into the 12-mm band of the
Tidbinbilla 70-m telescope. This gave 70 targets at $z>2.3$ out of the
878 PQFS sources, and limiting the sample further to sources with flux
densities in excess of $0.5$ Jy at $\approx0.4$ GHz, gave a total of
19 sources at $z>2.9$.



\section{Observations and results}
\label{obs}

\subsection{The redshifted decimetre wave observations}
\label{gobs}

The redshifted decimetre wave observations where performed with the
Giant Metrewave Radio Telescope (GMRT)\footnote{The GMRT is run by the
  National Centre for Radio Astrophysics of the Tata Institute of
  Fundamental Research.} in March 2004, during the first run as
described in Paper I. Again, as per Paper~I, we searched for
absorption within the host at the emission redshift of the quasar. Our
target sample therefore consisted (primarily) of objects in the PQFS
where either the \HI\ 21-cm or OH 18-cm ($^{2}\Pi_{3/2} J = 3/2$)
transition was redshifted into the 90-cm ($\approx327$ MHz) band. Due
to time constraints during the observations, we prioritised the
targets in which $B\gapp19$, in order to preferentially select sources
where the presence of dust, associated with dense gas, would dim the
visible and ultra-violet light. This left 13 out of the original 19
targets (see Table~\ref{sum}).

\begin{figure*}
	\vspace{23cm} \setlength{\unitlength}{1in}
\includegraphics{0335-122.ps}
\includegraphics{0347-211.ps}
\includegraphics{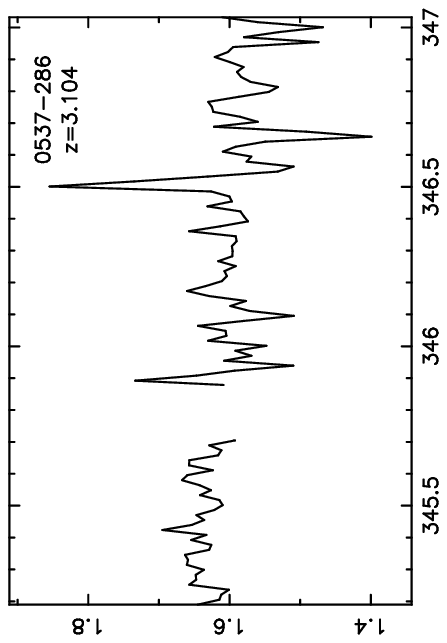}
\includegraphics{1026-084.ps}
\includegraphics{1228-113-HI.ps}
\includegraphics{1228-113-OH.ps}
\includegraphics{1251-407.ps}
\includegraphics{1351-018-HI-2.ps}
\includegraphics{1351-018-OH.ps}
\includegraphics{1535+004-HI.ps}
\includegraphics{1535+004-OH.ps}
\includegraphics{1937-101-H1_1a.ps}
\includegraphics{1937-OH-final.ps}
\includegraphics{2215+02-HI.ps}
\includegraphics{2215+02-OH.ps}
\caption{The GMRT spectra. The ordinate in each
spectrum shows the flux density [Jy] and the abscissa the Doppler
corrected frequency [MHz]. All spectra have been extracted from the
spectral cube, with the exception 1351--018 at 302 MHz and 2215+02 at
365 MHz, where the visibilities are averaged together. For the latter and
2215+02 at 311 MHz the flux scale is not absolute. }
\label{spectra}
\end{figure*}

We used all 30 antennae and the 90-cm receiver over a 2 MHz bandwidth
(giving a coverage of $\approx\pm900$ \kms). Over the 128 channels (2
polarisations) this gave a channel width of 16 kHz, or a velocity
resolution of $\approx15$ \kms. For all of the runs we used 3C\,48,
3C\,147 and 3C\,286 for bandpass calibration and used separate phase
calibrators for all of the sources, as heavy flagging of the target
data could result in poor self calibration.  The flagging and all of
the reduction was done by the {\sc miriad} interferometry reduction
package.  Synthesised beam sizes were typically $\gapp10''$ for the
90-cm and $\approx6''$ for the 21-cm observations (OH $^{2}\Pi_{1/2} J
= 1/2$ redshifted from 6-cm).  As per the other observations
(Paper~I), channel 117 of the RR polarisation and the telescope
pairing between antennae E02 and E03 (15 and 16 in {\sc aips}/{\sc
miriad} convention), were removed. Phase stability on all but the
shortest baselines was excellent.  In this band radio frequency
interference (RFI) was considerably more severe than at 21-cm \& 50-cm
(Paper~I). Regarding each source:\\ 0335-122 was observed for 1.5
hours at a centre frequency of 319.77 MHz.  While there were no
overwhelmingly bad frequencies, after flagging inferior data only 135
good antenna pairs were retained.\\ 0347--211 was observed for 1.4
hours at a centre frequency of 360.14 MHz. This band was relatively
RFI free and 300 good antenna pairs were retained.\\ 0537--286 was
observed for 1.4 hours at a centre frequency of 346.10 MHz, with 210
good antenna pairs. High amplitude spikes from 345.7 to 345.9 MHz,
required the removal of these frequencies.\\ 0913+003 was observed for
1.4 hours at a centre frequency of 348.65 MHz. A spike was also
dominant over 348.1 to 348.3 MHz and the full band before 18:00 UT was
swamped with RFI, leaving only 40 minutes on source over a fragmented
band (347.5 to 347.9 MHz). Also, since the bandpass was observed at
17:00 UT, no gain calibration is possible for this source.\\ 1026--084
was observed for 1.9 hours at a centre frequency of 316.03 MHz and 250
good antenna pairs were retained. Again severe RFI was present over
315.3 to 315.6 MHz and frequencies above $316.7$ MHz.\\ 
1228--113 was searched for \HI\ in a 1.4 hour observation centered on 313.69
MHz. In the 210 good antenna pairs severe RFI remained from 313.0 to
313.3 MHz, leaving little remaining band below these frequencies.
Since the OH $^{2}\Pi_{1/2} J = 1/2$ (4751 MHz) transition at the
redshift of this source fell into the 20-cm band of the GMRT, this
line was searched at a centre frequency of 1049 MHz for 4.4 hours. RFI
was minimal and 350 good antenna pairs were retained.\\ 1251--407 was
observed for 1.4 hours at a centre frequency of 305.38 MHz. This band
was fairly clean and 285 good antenna pairings were retained.\\ 
1351--018 was searched for \HI\ in a 1.0 hour observation centered on
301.76 MHz.  Although there was no overwhelmingly apparent RFI over
300.8 to 302.7 MHz, the reference antenna was affected, and so no
reliable image could be produced. The optical depth is therefore
derived from the averaged visibilities of the 260 good antenna
pairings. OH was also searched at a frequency of 354.23 MHz for 1.4
hours. No particular frequency was especially subject to RFI, although
flagging of some affected baselines was required, leaving 310 good
antenna pairs. \\ 
1535+004 was searched for \HI\ in a 1.4 hour
observation centered on 315.86 MHz.  Very little RFI was present
giving excellent data over the 400 good antenna pairs used. Again,
since the OH $^{2}\Pi_{1/2} J = 1/2$ (4751 MHz) transition at the
redshift of this source fell into the 20-cm band of the GMRT, this
transition was searched at a centre frequency of 1056 MHz for 3.1
hours. One polarisation (LL) was severely affected by RFI, and
subsequently removed, while for the remaining polarisation RFI was
minimal and 370 good antenna pairs were retained.\\ 
1630--004 was
observed for 1.0 hour at a centre frequency of 321.07
MHz. Unfortunately RFI totally swamped this band.\\ 
1937--101 was searched for \HI\ in a 1.4 hour observation centered on 296.72 MHz. No
major RFI was present and 300 good antenna pairs were retained.
OH was also searched at a frequency of 348.31 MHz for 1.8 hours. RFI was
minimal and 370 good antenna pairs where retained. \\
2215+02 was searched for \HI\ in a 1.4 hour observation centered on
310.67 MHz. Although no particular frequencies where affected by RFI,
only 190 antenna pairs proved to be of good quality.
OH was also
searched at a frequency of 364.69 MHz for 3.54 hours. Due to RFI many
baseline pairs had to be removed, leaving 210 pairs. Note that the
bandpass calibrator, 3C\,295, used for this source is unknown to {\sc
miriad} and so this could not be used to correct the gains for
2215+02. The flux density scale for this source (Fig. \ref{spectra})
should therefore not be considered as absolute. Furthermore, unlike
the lower frequency \HI\ observations, the phase stability was very
poor during this observation, preventing the extraction of a high
quality cube. \\ 2245--059 was observed for 1.4 hours at a centre
frequency of 330.71 MHz.  This band was so dominated by RFI, that no
useful data could be retained.
	
\subsection{The redshifted millimetre wave observations}
 \label{tobs}

The redshifted millimetre wave observations were performed with the
Australia Telescope's Tidbinbilla 70-m telescope\footnote{The
  Australia Telescope is funded by the Commonwealth of Australia for
  operations as a National Facility managed by CSIRO.}, over several
sessions between November 2003 and March 2005. Again, the sources were
selected according to those in which a strong transition\footnote{That
  is, transitions of CO, HCO$^+$ and HCN, which are optically thick in
  the four known redshifted millimetre absorption systems.} fell into
the 1-cm ($\approx22$ GHz) band.  For the backend we used the
das\_xxyy\_64\_2048 configuration (dual polarisation with 2048
channels over a 64 MHz band), giving a resolution of $\approx0.5$
\kms\ over a range of $\approx\pm500$ \kms: Such high resolution was
required, since although optically thick, the line-widths in the four
known millimetre absorption only span a few
\kms\ \citep{wc94,wc95,wc96b,wc98}, cf. up to $240$ \kms\ for the,
optically thin, OH absorption (see figure 5 of \citealt{cdbw07}).
\begin{figure*}
\vspace{8.5cm}
\includegraphics{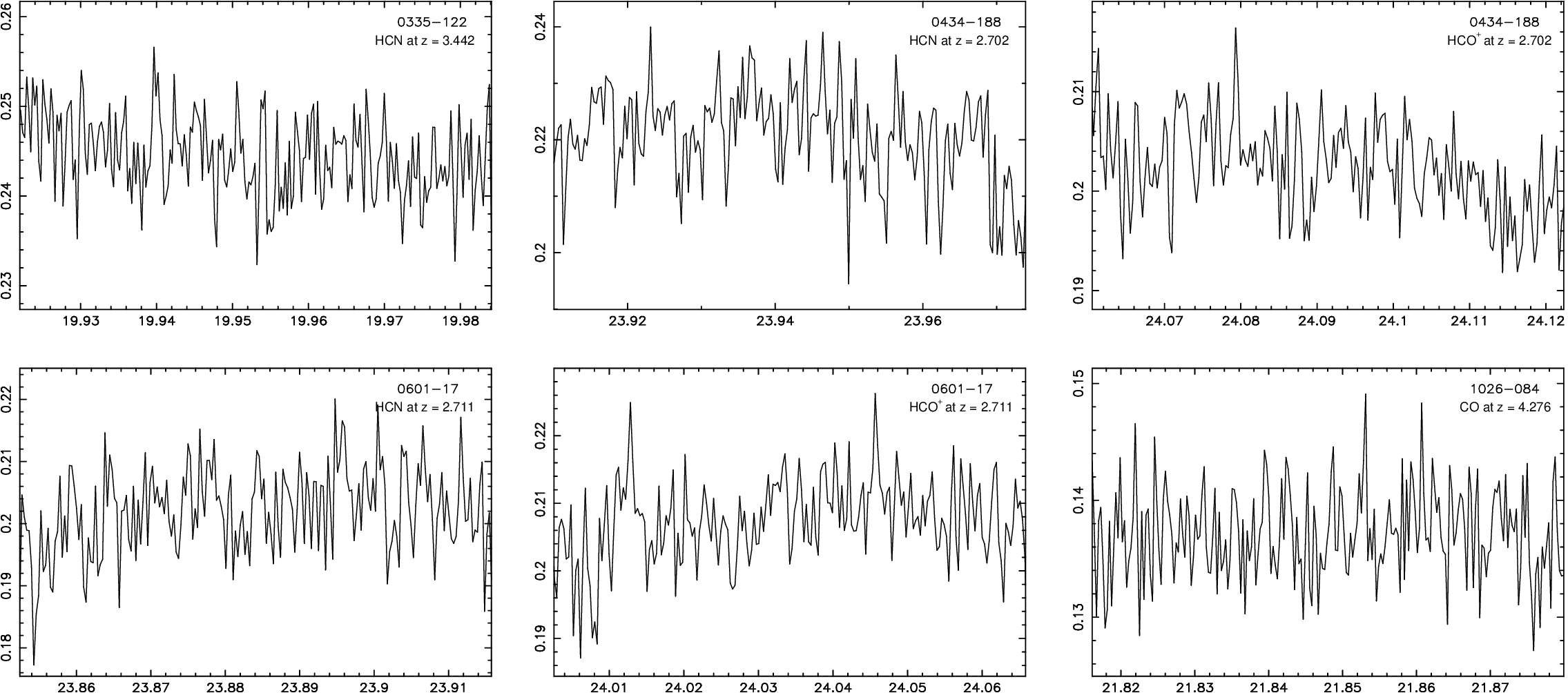}
\caption{The Tidbinbilla spectra. The ordinate in each spectrum shows
the flux density [Jy] and the abscissa the Doppler corrected frequency
[GHz]. Each spectrum has been smoothed to a channel width of 4 \kms.}
\label{tid-spectra}
\end{figure*}

The half power beam-width (HPBW) of the single dish at these
frequencies is $\approx50''$ and, in order to ensure good baselines,
we used position switching with 1 minute per position over a 4-point
position switching cycle, each source being observed for a total of 4
hours. In general, the baselines were excellent with no baseline
subtraction of the spectra (Fig. \ref{tid-spectra}) being required.
Flagging of data was only required when communication with the
sub-reflector had been lost. During the observations, the $T_{\rm
  A}^{\ast}$ flux scale was periodically calibrated against a thermal
load provided by a noise diode. An additional correction for the
variation of antenna gain with elevation was applied to each spectrum
off-line. The data were reduced using the {\sc dfm}\footnote{Data From
  Mopra, available from
  http://www.phys.unsw.edu.au/astro/mopra/software.php}, graphical
interface to the {\sc
  spc}\footnote{http://www.atnf.csiro.edu.au/software} package. A
quotient was formed between the source and reference positions and the
resulting spectra were averaged and weighted by $T_{\rm int}/T_{\rm
  sys}^2$.  Lastly, the data were corrected onto the $T_{\rm MB}$
temperature scale by multiplying by the beam efficiency at 22\,GHz
($\eta \approx$0.48, \citealt{gbe+03}).

\subsection{Results}
\label{results}

In Table \ref{sum} we summarise our results. The column density is
derived from the velocity integrated depth,
$\tau\equiv-\ln\left(1-\frac{\sigma}{f\,S}\right)$, where $\sigma$ is
r.m.s. noise in the case of our exclusive non-detections and $S$ and $f$ the
flux density and covering factor of the background continuum source,
respectively. In the optically thin regime (where $\sigma/f.S\lapp0.3$),
which applies to the vast majority of the known \HI\ and all
of the OH absorbers, the $1\sigma$ column density limit is given by
\begin{equation}
N = X\times \frac{T}{f}\int\!\frac {\sigma}{S}\,dv,
\label{equ1}
\end{equation}
where for \HI\ 21-cm, $X=1.823\times10^{18}$ \citep{wb75} and $T$ is
the spin temperature ($T_{\rm s}$) and for OH, $X=2.38\times10^{14}$
for the $^{2}\Pi_{3/2} J = 3/2$ (1667 MHz) transition and
$X=1.61\times10^{14}$ for the $^{2}\Pi_{1/2} J = 1/2$ (4751 MHz)
transition \citep{hgb87}, with $T$ being the excitation temperature
($T_{\rm x}$).

For OH, the values of $X$ are derived from the total column
density obtained from a rotational transition, which is given by
\begin{equation}
 N = \frac{8\pi}{c^3}\frac{\nu^{3}}{g_{J+1}A_{J+1\rightarrow J}}\frac{Qe^{E_J/kT_{\rm x}}}{1-e^{-h\nu/kT_{\rm x}}}
\left.\int\right.\tau dv,
\label{equ2}
\end{equation}
where $\nu$ is the rest frequency of the $J\rightarrow J+1$
transition, $g_{J+1}$ and $A_{J+1\rightarrow J}$ are the statistical
weight and the Einstein A-coefficient\footnote{These are taken from
  \citet{cklh95,cms96} or derived from the dipole moment
  (e.g. \citealt{rw00}).} of the transition, respectively, and $Q =
\sum^{\infty}_{J=0}g_{J}~e^{-E_J/kT_{\rm x}}$, with $g_{J} = 2J+1$, is
the partition function\footnote{The energy of each level, $E_J$, is
  obtained from the JPL Spectral Line Catalog \citep{ppc+98}. An
  on-line column density calculator based on Equ. \ref{equ2} is
  available at http://www.phys.unsw.edu.au/$\sim$sjc/column/}.  Since
$kT_{\rm x}\gg h\nu$ for $T_{\rm x}\approx10$ K and $\nu\lapp5$ GHz,
Equ. \ref{equ2} can be simplified to the above expression
(Equ. \ref{equ1})\footnote{Where the values for $X$ are derived using
  a weighting factor of $\sum g_{J}/g_{J} = 16/5$ and
  $A_{J+1\rightarrow J}=7.71\times10^{-11}$~s$^{-1}$ \citep{cm67} for
  the 1667 MHz transition. For the 4751 MHz transition, $\sum
  g_{J}/g_{J} = 8/3$ and $A_{J+1\rightarrow
    J}=7.7597\times10^{-10}$~s$^{-1}$
  (http://www.strw.leidenuniv.nl/$\sim$moldata/datafiles/oh@hfs.dat).}.
For the millimetre-wave column densities, based on the four known
systems, the covering factor is expected to be close to unity
\citep{wc94,wc95,wc96b,wc98} so, as in the optically thin regime, this
can be written outside of the integral. However, for the same excitation
temperatures the higher frequencies give $kT_{\rm x}\sim
E_J\sim h\nu$ and so the column density cannot be approximated via a
linear dependence on the excitation temperature. We
therefore adopt the canonical value of 10 K at $z=0$\footnote{Since we
  are analysing high redshift sources, we scale this with the
  temperature of the cosmic microwave background, $T_{\rm CMB} = 2.73
  (1 + z)$, giving $T_{\rm x}=17 - 21$ K over $z=2.702 - 4.276$.}.
\begin{table*} 
\centering
\begin{minipage}{180mm}
\caption{Summary of the search for decimetre absorption lines in the hosts of
$z\gapp3$ PQFS sources.  $\nu_{\rm obs}$ is the observed frequency of
the line [MHz], $\sigma_{{\rm rms}}$ is the r.m.s. noise [mJy] reached
per $\Delta v$ channel [\kms], $S_{\rm cont}$ is the continuum flux
density (uncalibrated for 2215+02, Sect. \ref{obs}) [Jy], $\tau$ is the
optical depth of the line calculated per channel, where
$\tau=3\sigma_{{\rm rms}}/S_{\rm cont}$ is quoted for these
non-detections, $N$ is the resulting column density [\scm], where $T_{\rm s}$
is the spin temperature of the \HI~21-cm line, $T_{\rm x}$ is the
excitation temperature of the corresponding OH line and $f$ the
respective covering factor. In all cases OH refers to the
$^{2}\Pi_{3/2} J = 3/2$ (1667 MHz) transition, with the exception of
1228--113 and 1534+004 for which we observed the $^{2}\Pi_{1/2} J =
1/2$ (4751 MHz) transition. Finally, in light of the results of Paper
I, we list the $V$, $B$ \& $K$ magnitudes (where available) with their respective
references.}
\begin{tabular}{@{}l l c c c c r c  c c c c c c @{}} 
\hline
PKS &  $z_{\rm em}$ & Line & $\nu_{\rm obs}$ &$\sigma_{{\rm rms}}$ & $\Delta v$ &$S_{\rm cont}$ & $\tau$ & $N$ & $z$-range & $B$ & $V$ &$K$ & Ref \\
\hline
0335--122 &  3.442 & \HI & 319.77& 23& 15& 0.90&$<0.077$ & $<2.1\times10^{18}.\,(T_{\rm s}/f)$& 3.428--3.456& 21.018 & 20.110& 17.510 & 1,2\\
0347--211 &2.944  &   \HI &360.14& 17 & 13 &0.51 & $<0.10$ &$<2.4\times10^{18}.\,(T_{\rm s}/f)$&2.935--2.953  & 20.476 &-- & 17.900 & 2\\
0537--286& 3.104 &  \HI &346.10 & 54& 14 & 1.61&$<0.10$ &$<2.6\times10^{18}.\,(T_{\rm s}/f)$ & 3.092--3.106 & 19.290 & -- & 16.770 &  2,3\\
...&...&...& ... &\multicolumn{5}{c}{RFI required removal of $z=3.1064-3.1088$} &  3.109--3.115 &...&...&...\\
0913+003 & 3.074 & \HI &348.65& --& 13& -- &\multicolumn{3}{c}{RFI dominant} & 20.998 & 20.775  &-- & 2,4 \\
1026--084 &4.276  & OH  & 316.03&53  & 15 &0.65 &  $<0.24$& $<1.5\times10^{15}.\,(T_{\rm x}/f)$& 4.265--4.283   & 21.070 & -- & -- & 2 \\
1228--113 &3.528  & \HI &313.69& 26&15 &0.49  & $<0.16$& $<4.3\times10^{18}.\,(T_{\rm s}/f)$&3.516--3.532  &22.010 &--  & 16.370 & 5,3 \\
...& ...&OH & 1049.17 &3.7& 18& 0.64& $<0.017$& $<4.9\times10^{13}.\,(T_{\rm x}/f)$&3.513--3.543  &... &... &  ...\\
1251--407 & 4.460& OH  & 305.38 & 12 & 15& 0.25 &$<0.14$&$<5.0\times10^{14}.\,(T_{\rm x}/f)$&
4.442--4.478  &   -- & 23.7& -- & 6\\
1351--018 & 3.707 & \HI& 301.76& --& 16& --& $<0.090$ &$<2.6\times10^{18}.\,(T_{\rm s}/f)$ &3.693--3.722   & 21.030 & 19.696& 17.070& 7,4,3 \\
...& ...&OH &354.23& 39&  13& 0.75&  $<0.052$&$<1.6\times10^{14}.\,(T_{\rm x}/f)$ &3.695--3.719 & ...& ...\\
1535+004 &  3.497 &\HI & 315.86& 8.6& 15& 0.37&  $<0.070$&$<1.9\times10^{18}.\,(T_{\rm s}/f)$  &3.485--3.509   & 22.500& --& 19.54 & 3 \\
...& ...&OH   &1056.41  & 4.0 & 18 & 0.47 & $<0.032$&$<9.4\times10^{13}.\,(T_{\rm x}/f)$ &3.486--3.514& ...& ...\\
1630--004&3.424  & \HI& 321.07& -- & 15& -- &\multicolumn{3}{c}{RFI dominant} & --& 21.81 &-- & 8 \\
1937--101&  3.787 & \HI& 296.72& 9.4 &16 &0.61 & $<0.046$ &  $<1.3\times10^{18}.\,(T_{\rm s}/f)$ &3.773--3.802 & 18.800 & --& 13.816 & 2,9\\
...& ...&OH  & 348.31 &3.5& 14& 0.51& $<0.021$& $<7.0\times10^{13}.\,(T_{\rm x}/f)$ & 3.775--3.795& ...& ...&...\\
2215+02	 &3.572  & \HI &310.67 & 14& 15 & 3.07& $<0.014$& $<3.7\times10^{17}.\,(T_{\rm s}/f)$ &3.558--3.585  & 21.840 &20.420& 19.340 & 10\\
...& ...&OH  &364.69 & 22 &13 & -- & $<0.089$&$<2.8\times10^{14}.\,(T_{\rm x}/f)$ &3.561--3.583 & ...&...&...&...\\
2245--059&3.295  & \HI& 330.71& --& 14&-- & \multicolumn{3}{c}{RFI dominant} & 19.523 & -- & --& 2\\
\hline       
\end{tabular}
{\flushleft References: (1) \citet{ehl05}, (2) SuperCOSMOS Sky Survey,
  \citet{hmr+01}, (3) P. Francis \citetext{priv.comm}, (4) SDSS DR6,
  \citet{aaa+08}, (5) \citet{cgkt06}, (6) \citet{jws+02}, (7) \citet{dwf+97}, (8) \citet{wmh+02}, (9) \citet{sh89}, (10) \citet{fww00}.}
\label{sum}  
\end{minipage}
\end{table*} 

\begin{table*} 
\centering
\begin{minipage}{132mm}
\caption{Summary of the search for millimetre absorption lines in the
hosts of $z\gapp3$ PQFS sources. $\nu_{\rm obs}$ is the observed
frequency of the line [GHz], $\sigma_{{\rm rms}}$ is the r.m.s. noise
[mJy] reached per 4 \kms\ channel, $S_{\rm cont}$ is the continuum
flux density [Jy] and $\tau$ is the optical depth of the line calculated
per channel. The column density [\scm] of each molecule is calculated for
$f=1$ and an excitation temperature of $T_{\rm x}=10$ K at $z=0$.
The magnitudes, where available, are given in Table \ref{sum}. For
0434--188, $B=20.86$ \citep{ehl05} \& $K=16.21$ \citep{dwf+97} and for
0601--17, $B=20.445$ \& $R=20.172$ \citep{hmr+01}.}
\begin{tabular}{@{}l l c c c c c c c  @{}} 
\hline
PKS &  $z_{\rm em}$ & Line & $\nu_{\rm obs}$ &$\sigma_{{\rm rms}}$ &$S_{\rm cont}$ & $\tau$ & $N$ & $z$-range \\
\hline
0335--122 &  3.442 & HCN $0\rightarrow1$  & 19.953 &  4.5 & 0.245 & $<0.055$ & $<9.7\times10^{12}/f$ &  3.435--3.449 \\
0434--188 & 2.702 & HCN $0\rightarrow1$  & 23.942 & 8.7& 0.220 & $<0.12$ & $<1.7\times10^{13}/f$ & 2.698--2.707 \\
...& ... & HCO$^+$ $0\rightarrow1$  & 24.092 & 4.6 & 0.202 &$<0.068$  & $<4.3\times10^{12}/f$ &2.698--2.707 \\
0601--17 & 2.711 & HCN $0\rightarrow1$  &23.884 & 7.0 & 0.202 & $<0.10$ & $<1.4\times10^{13}/f$ &2.706--2.716 \\
...& ... & HCO$^+$ $0\rightarrow1$  & 24.034 & 6.3 & 0.207 &$<0.091$ &  $<5.7\times10^{12}/f$ &2.705--2.716 \\
1026--084 &  4.276 & CO $0\rightarrow1$  &21.848 & 3.9 & 0.137 &  $<0.085$& $<7.8\times10^{15}/f$&  4.268--4.284\\
\hline       
\end{tabular}
\label{tid-sum}  
\end{minipage}
\end{table*} 


\section{Possible effects in the non-detection of atomic absorption}
\label{penaa}

From Table \ref{sum}, it is seen that we do not detect 21-cm
absorption in any of the ten sources for which good data were
obtained. In our earlier study of sources from the PHFS (Paper I), we
detected 21-cm absorption in one out of four sources, although the
only real difference between the PHFS and PQFS samples is that the
PQFS has a lower flux limit (0.25 Jy, cf.\ 0.5 Jy)\footnote{Also a
slightly stricter lower limit to the spectral index ($\alpha>-0.4$,
cf. $\alpha>-0.5$, where $S_\nu\propto\nu^\alpha$).}. However, due to
our restriction that the sources are at redshifts where the HCO$^+$
$0\rr1$ transition is redshifted into the 12-mm band, unlike the
previous search which spanned $z\approx0.1 - 3.5$, our targets are all
at $z\gapp2.9$, where published searches have been very rare
(Fig. \ref{tau-z}):
\begin{figure}
\vspace{6.8cm}
\includegraphics{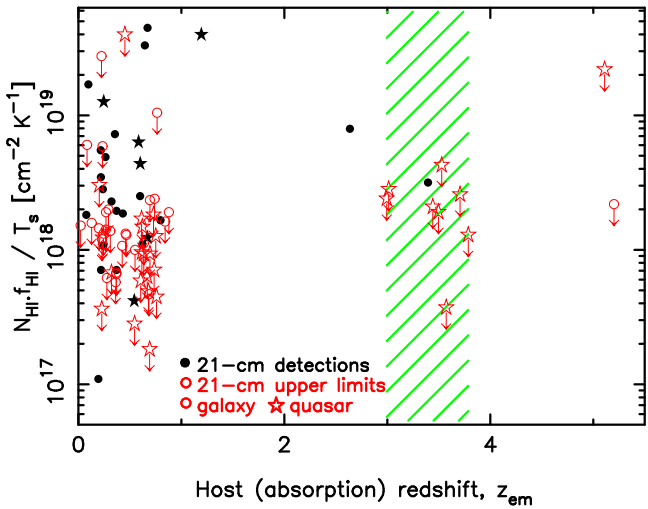}
\caption{The scaled velocity integrated optical depth of the \HI\ line
  ($1.823\times10^{18}.\int \tau dv$) versus the host redshift for the
  published searches for associated 21-cm absorption (see Table
  \ref{t1} and Appendix A). The filled symbols represent the 21-cm
  detections and the unfilled symbols the non-detections, with stars
  designating quasars and circles galaxies (see Sect. \ref{sec:class}). The 
  hatched region shows the range of our \HI\ searches
  (Table~\ref{sum}). The other results are from the references quoted
  in Table~\ref{t1} (see Appendix A), with the addition of the two
  $z_{\rm em}>5$ non-detections of \citet{cwh+07}.}
\label{tau-z}
\end{figure}
In the figure all of the sources in the hatched region are from our
search (Table~\ref{sum}), in addition to the $z_{\rm em}=3.497$ quasar
1535+004 from Paper~I. However, there is one 21-cm detection in this
band ($z=3.4$ in 0902+343, \citealt{ubc91}) and so the high redshift
selection on its own cannot be responsible for our non-detections, all
of which have been searched to limits comparable to the detections.

Since we are searching for associated absorption, although covering
factors may be an issue, these will not be subject to the same
geometrical effects found for intervening systems: Specifically damped
Lyman-$\alpha$ absorption systems (DLAs), where \citet{cw06}
show a strong correlation between low absorber/quasar angular diameter
distance ratios and 21-cm detections, while high ratios are correlated
with non-detections. This suggests that the absorbers located
``closer'' to us (in an angular sense) have significantly larger
covering factors than those which share a similar angular diameter
distance to the background quasar.

Therefore our lack of 21-cm detections must be due to another effect
-- either low neutral hydrogen column densities, high spin
temperatures or low covering factors (see Equ.~\ref{equ1}), although,
as stated, the latter would have to be due to small intrinsic
absorption cross sections, since for both the 21-cm detections and
non-detections in associated systems, the absorber/quasar angular
diameter distance ratios are close to unity.  In the absence of
measurements of the neutral hydrogen column densities from the
Lyman-\AL\ line, unlike DLAs, we cannot compare how $N_{\rm HI}$
varies between the detections and non-detections nor speculate on spin
temperature effects, although \citet{cmp+03,ctp+07} have shown that
spin temperatures in DLAs may not vary by nearly as much as was
previously believed \citep{kc02}, and that these may generally be
below $T_{\rm s}\approx2000$ K (at least up to $z_{\rm
  abs}\approx3.4$).  Again, however, for these sources we have a
degeneracy of three variables, all of which may be mutually coupled
(see \citealt{ctp+07}), thus making the relative contribution of each
very difficult to ascertain.


\subsection{Incident fluxes}

\subsubsection{21-cm luminosities}
\label{21-cm fluxes}

Although our sample could be highly heterogeneous, since our only
criterion for our own targets is that they are high redshift PQFS
sources (Sect. \ref{intro}), there must be some reason why 21-cm
absorption is detected in some quasars and radio galaxies, while not
being detected in others, particularly at high redshift. \citet{cw06}
previously investigated a ``proximity effect'' in DLAs, where a high
21-cm flux may maintain a high population in the upper hyperfine
level, thus having relatively few anti-parallel spin atoms available
to absorb the 1420 MHz radiation \citep{wb75}. Although there were a
few non-detections and zero detections at incident flux densities of
$\gapp10^4$ Jy in this sample, there was no overwhelming trend, with
the aforementioned geometrical effects being apparently much more
significant with regard to the detection of 21-m absorption.

When we consider our sample, we cannot determine incident fluxes for
  the non-detections, as we have no knowledge of where the neutral gas
  would be located relative to the emission region. However, since we
  are just looking for statistical differences in this sample, we can
  still investigate this effect through the luminosities. The
  specific luminosity of the quasar at the rest frame emission
  frequency, $\nu_{\rm em}$, is $L_{\nu}=4\pi \, D_{\rm QSO}^2
  \,S_{\rm obs}/(z_{\rm em}+1)$, where $D_{\rm QSO}$ is the luminosity
  distance to the quasar\footnote{Throughout this paper we use
  $H_{0}=75$~km~s$^{-1}$~Mpc$^{-1}$, $\Omega_{\rm matter}=0.27$ and
  $\Omega_{\Lambda}=0.73$.}, $S_{\rm obs}$ is the observed flux
  density\footnote{Since we are searching for absorption at the host
  redshift, the quasar frame 21-cm flux density is given by the
  observed value.} and $z_{\rm em}+1$ corrects for the redshifting of
  the frequency increment.
\begin{figure*} 
\centering
\includegraphics[angle=270,scale=0.70]{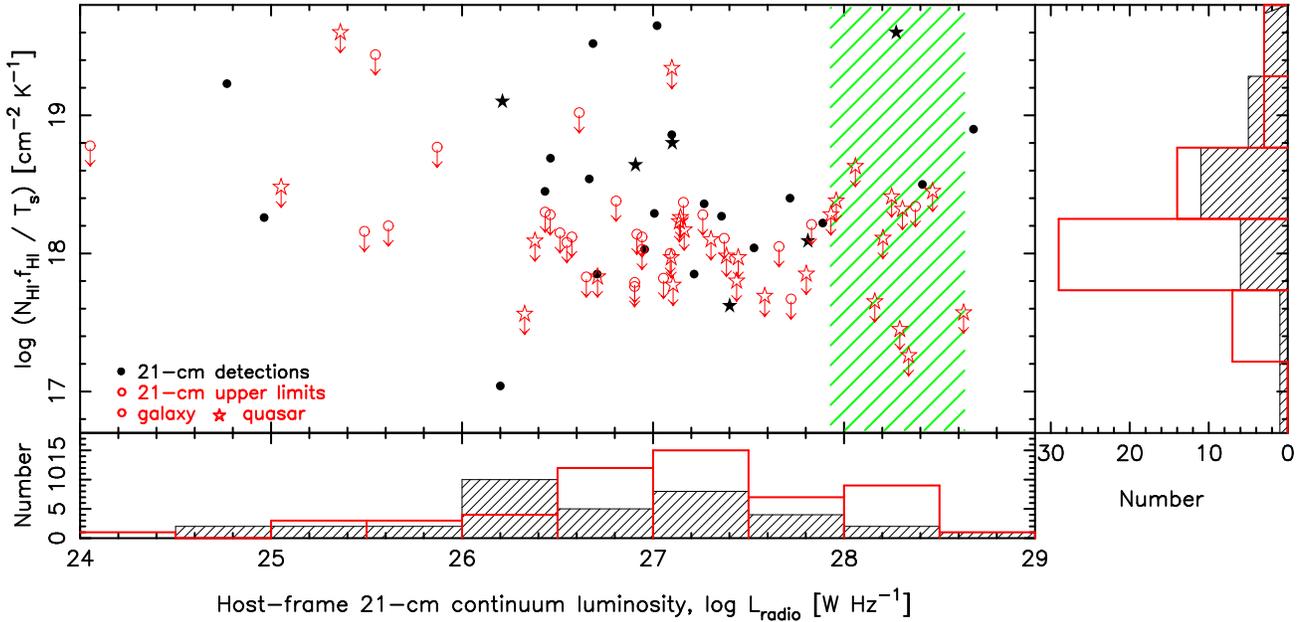}
\caption{The scaled velocity integrated optical depth of the \HI\ line
versus the quasar frame 21-cm luminosity for the quasars searched for
associated \HI\ absorption. Again, the hatched region
shows the range of our \HI\ searches. Throughout this paper the filled
symbols/hatched histogram represent the 21-cm detections and the unfilled
symbols/coloured histogram the non-detections.}
\label{lum}
\end{figure*}
Using this expression, in Fig. \ref{lum} we show the derived quasar
frame luminosities in relation to the velocity integrated optical
depth of the 21-cm absorption.  From this, the vertical histograms
show that, on the whole, the non-detections have been searched
sufficiently deeply, many to a higher sensitivity required for the
detections. The horizontal histograms show considerable overlap, and,
while the average $\log L_{\rm radio}$ is higher for the
non-detections, the difference is small ($\log L_{\rm radio}=27.06$
cf.\ $26.71$), and the distributions of luminosities for detections
and non-detections are not statistically different (the probability of
the null hypothesis for the Kolmogorov-Smirnov test is
33.7\%). Therefore, although a high incident 21-cm flux may make a
detection less likely through a highly populated upper hyperfine
level, this does not appear to be an overwhelming cause of the
non-detections.

\subsubsection{Ultra-violet luminosities}
\label{uvf}

Although a 21-cm proximity effect is not apparent for DLAs or
associated absorbers, in the ultra-violet such an effect is well
known, where a high ionising flux from the quasar is believed to be
responsible for the decrease in the number density of the ultra-violet
Lyman-\AL\ lines as $z_{\rm abs}$ approaches $z_{\rm em}$
\citep{wcs81,bdo88}. To excite the hydrogen beyond the realm of 21-cm
absorption does not require ionisation of the gas (by 912
\AA\ photons), but ``merely'' excitation above the ground state by a
Lyman-\AL\ (1216 \AA) photon, although the lifetime in this state is
only $\sim10^{-8}$~s. In any case, since both the ionising and Lyman-\AL\ photons
are $\sim10^{6}$ times as energetic as the spin-flip transition, if
the gas is excited by Lyman-\AL\ absorption, much of it will also be
ionised.

\begin{figure*}
\centering
\includegraphics[angle=270,scale=0.70]{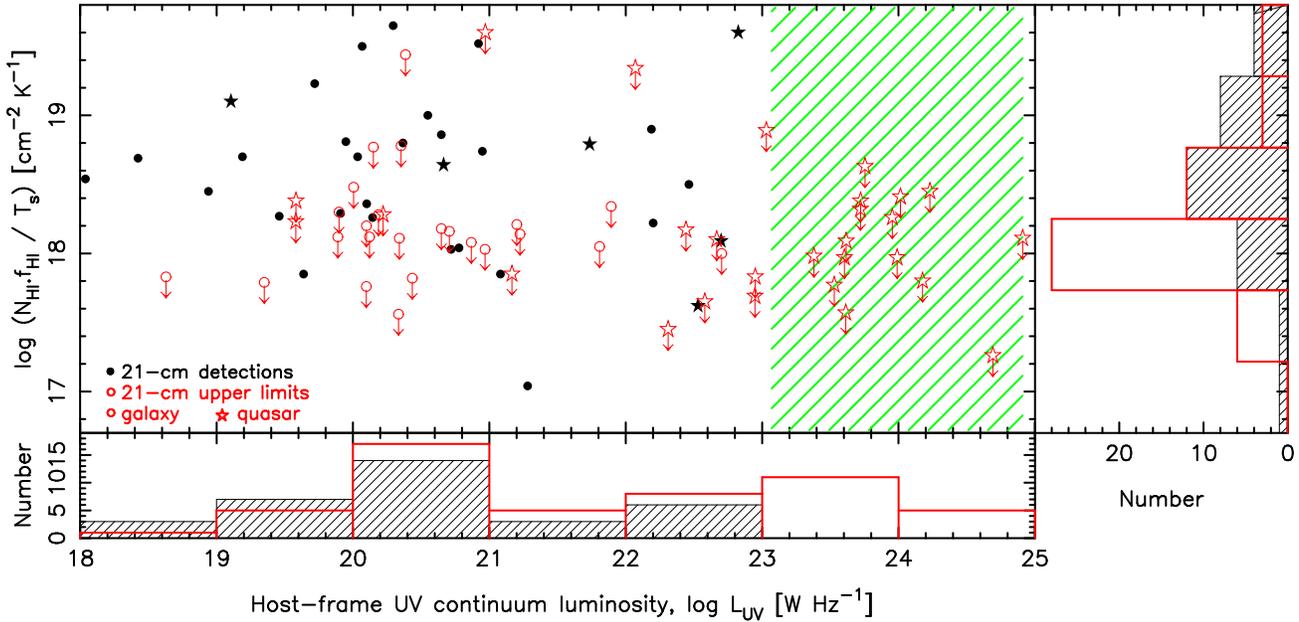}
\caption{As per Fig. \ref{lum}, but with the scaled velocity
integrated optical depth of the \HI\ line versus the quasar frame
ultra-violet luminosity.}
\label{ion}
\end{figure*}
Therefore, in order to determine the $\sim1000$ \AA\ fluxes, we have
exhaustively searched the literature and on-line archives to obtain
optical and near-infrared photometry of as much of the sample as
possible (see Appendix A). In some cases, we use the photometry of the
Sloan Digital Sky Survey \citep[SDSS,][]{yaa+00}\footnote{The SDSS is managed by the Astrophysical Research Consortium (ARC) for
the Participating Institutions. The Participating Institutions are the
American Museum of Natural History, Astrophysical Institute Potsdam,
University of Basel, University of Cambridge, Case Western Reserve
University, The University of Chicago, Drexel University, Fermilab,
the Institute for Advanced Study, the Japan Participation Group, The
Johns Hopkins University, the Joint Institute for Nuclear
Astrophysics, the Kavli Institute for Particle Astrophysics and
Cosmology, the Korean Scientist Group, the Chinese Academy of Sciences
(LAMOST), Los Alamos National Laboratory, the Max-Planck-Institute for
Astronomy (MPIA), the Max-Planck-Institute for Astrophysics (MPA), New
Mexico State University, Ohio State University, University of
Pittsburgh, University of Portsmouth, Princeton University, the United
States Naval Observatory, and the University of Washington.}, taken from Data
Release 6 \citep{aaa+08}, applying the transformations of
\citet{fig+96} to ensure consistency between the bands. From these
measurements, we estimate the $\lambda\approx1216(1+z)$\AA\ flux
according to the prescription in Appendix~B, and
thence the luminosity in the rest-frame of the galaxy/quasar.

Since the sources cover a range of redshifts, the
interpolations/extrapolations necessary to estimate $L_{\rm UV}$
involve different bands for different sources. We can check the
validity of the estimates by comparing the interpolated value of the
flux at the nominal 1216\AA\ wavelength (from $BVR$) with that
extrapolated from the $JHK$ bands. The latter situation is similar to
the case of extrapolating from optical bands for low-redshift
sources. In our sample, there are only three sources in the high
($z>2.5$) redshift group for which we have more than one near-infrared
band available: J0414$+$0534, 1937$-$101 (both of which have
2MASS\footnote{The Two Micron All Sky Survey, is a joint project of
  the University of Massachusetts and the Infrared Processing and
  Analysis Center/California Institute of Technology, funded by the
  National Aeronautics and Space Administration and the National
  Science Foundation.}  photometry) and 2215$+$020 \citep{fww00} and
for these, we can compare the extrapolation of the $JH$ measurements
to that obtained from $BR$. 2215$+$020 has the same extrapolation for
both, 1937$-$101 has $JH$ overestimating the flux by a factor of
$\sim5$, and J0414$+$0534 overestimates by $\sim50$. The analysis of
the latter source could be affected by the extreme optical--near-IR
colour of $V-K=10.26$ \citep{lejt95} [see Sect. \ref{nomol}], most
likely due to the presence of dust in the intervening gravitational
lens or host galaxy\footnote{Note that on the basis of their detection
  of \HI\ in the lens in conjunction with a strong limit on the OH
  column density, \citet{cdbw07} [and references therein] suggest that
  the reddening is occurring in the $z=2.64$ host galaxy.}. The other
two sources, however, show that the extrapolation from longer
wavelengths (rest-frame optical) can give a value of the UV flux in
broad agreement with the extrapolation from rest-frame UV. We note
that an extrapolation using $JH$ for these sources is roughly
equivalent to using $BR$ for sources at $0.6\lapp z \lapp 1.0$.

Plotting our results, in Fig. \ref{ion} we see an equal mix of 21-cm
detections and non-detections below the median value of the UV
luminosity range $\left(\frac{18.04+24.91}{2} =
21.48\right)$. However, at higher luminosities the distribution is
dominated by non-detections which become exclusive at $L_{\rm
  UV}\gapp10^{23}$ W Hz$^{-1}$, the range occupied by our observed
sample (Table~\ref{obs})\footnote{The mean luminosity of the
  detections is $\log L_{\rm UV}=20.5\pm1.2$, cf. $\log L_{\rm
    UV}=21.8\pm1.7$ for {\em all} of the non-detections and $\log
  L_{\rm UV}=20.8\pm1.1$ for the low luminosity
  non-detections.}. Investigating the likelihood of such a
distribution occurring by chance, defining a partition at the median
of the sample gives 52 objects (26 detections \& 26 non-detections) in
the low luminosity bin and 33 objects (7 detections \& 26
non-detections) in the $\log L_{\rm UV} > 21.48$ bin. For an unbiased
sample, i.e. there is an equal likelihood of either a detection or
non-detection (as may be expected from orientation effects, see next
section), the binomial probability of 26 or more detections out of 52
objects occurring is 55\%. However, the probability of 25
non-detections or more out of 33 objects in the other bin is just
0.23\%. Moving the partition to $\log L_{\rm UV} = 23.0$, thus
covering the range of our targets (Fig. \ref{ion}), in the low
luminosity bin there are 33 detections out of 69 objects (again a near
50\% detection rate) and in the high luminosity bin, 0 detections out
of 16. Again, assuming an unbiased sample, the binomial probability of
this latter distribution is 0.0015\%, a significance of $4.3\sigma$
assuming Gaussian statistics\footnote{Should there be an overwhelming
  error in the band extrapolations, to the point where the $z>1$
  luminosities cannot be used in the comparison, for the $z<1$ sources
  {\em alone} the result remains significant at $2.9\sigma$.}, against
the probability of a 21-cm absorption detection being unrelated to the
ultra-violet luminosity. Applying a Kolmogorov-Smirnov test, there is
a 99.25\% confidence that the UV luminosity distribution of the
detections differs from that of the non-detections. Therefore there is
a relationship between the ultra-violet luminosity and the likelihood
of detecting 21-cm absorption.




\subsection{Orientation effects}
\label{oeff}

\subsubsection{Unified schemes of AGN}

Previously, the mix of 21-cm detections and non-detections has
been attributed to unified models of active galactic nuclei
(AGN). This is an orientation effect due to the presence of a dense
sub-pc circumnuclear disk or TORUS\footnote{e.g. Thick Obscuration Required by Unified
Schemes \citep{con98}.} of gas which obscures the UV/optical light
from the AGN: In type-1 objects, the obscuration has its rotation axis directed
towards us and the AGN and the centralised broad-line region are
viewed directly, whereas in type-2 objects the AGN is hidden, and only
the more extended narrow-line region is visible (see \citealt{ant93,up95}).
It is hypothesised that the 21-cm absorption occurs in this
obscuration and so is only visible in type-2 objects, where the gas
intercepts the line-of-sight to the AGN (e.g. \citealt{jm94,cb95}).

\begin{figure}
\vspace{6.8cm}
\includegraphics{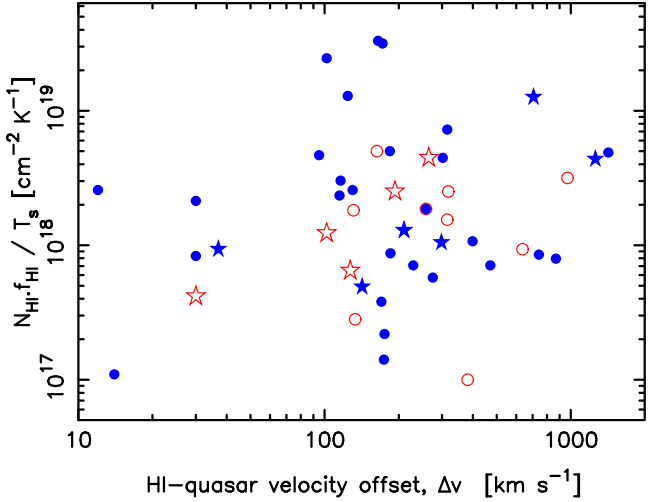}
\caption{The 21-cm absorption strength the versus the velocity offset
of the 21-cm absorption from the host. Here we have plotted each
individually resolved absorption component at its line strength and
velocity offset for each of the detections (Table \ref{t1}). The solid
symbols represent blueshifted (approaching) \HI\ and the hollow symbols
redshifted (receding) \HI. In associated systems these offsets
represent peculiar motions rather than the Hubble flow. Furthermore,
there are often velocity differences between different optical lines
and so these velocities cannot be reliably converted to distance
offsets to yield the incident flux on the absorbing gas.}
\label{tau-vel}
\end{figure}

For instance, of four 21-cm detections in a sample of 23 $z_{\rm
  em}<0.7$ radio galaxies \citep{mot+03}, two are narrow-line radio galaxies,
indicating type-2 objects, and the other two are weak-line radio
galaxies. Therefore, superficially at least, this is consistent with
the 21-cm absorption occurring in an intervening torus or disk
\citep{mot+03}. However, the fact that there are significant redshift
(velocity) offsets between the 21-cm and optical lines, leads the
authors to conclude that some of the \HI\ responsible for the
absorption is located farther out than the central sub-pc occupied by
the dense obscuration. The presence of significant gas motions, due to
infall, outflows, jet interactions as well as general galactic
rotation\footnote{Although gas in the galactic disk may be expected to
  share the same orientation as the sub-pc obscuration
  (\citealt{cjrb98}, cf. \citealt{gbe+03}, see also
  \citealt{cur99p}).}  is also suggested by the large velocity offsets
found in a sample of 19 detections by \citet{vpt+03} [see
  Fig.~\ref{tau-vel}].

Furthermore, \citet{pcv03} note that 21-cm absorption is more likely
to be detected in radio galaxies than in quasars, suggesting that the
quasars are viewed at lower inclinations, thus having the observed radio
emission bypass the absorbing gas which obscures the optical/UV
light. The orientations for a very limited sample \citep{pcv03} are quantified by
the core prominence, where the galaxies have core emission fractions
of $f_{\rm c}<0.03$, cf. $f_{\rm c}>0.04$ for the quasars, the larger core
prominence being attributed to a more direct viewing angle to the AGN.
From a larger sample, \citet{gs06a} find that the galaxies have a
median value of $f_{\rm c}=0.011$, compared to $f_{\rm c}=0.028$ for
the quasars. However, the distribution becomes more mixed with nearly
half of the quasars located below the median value for the galaxies ($f_{\rm
c}\sim0.01$) and with five galaxies at $f_{\rm
c}\gapp0.03$. Additionally, although none of the quasars have been
detected in 21-cm absorption, all of these five galaxies exhibit
strong ($N_{\rm HI}\geq1.6\times10^{18}.\,T_{\rm s}/f$ \scm)
absorption, thus weakening the argument that core dominance indicates
lower inclinations, or at least, the amount of cold neutral gas lying
along the sight-line to the AGN.

Many of these core fractions have however been
estimated\footnote{Using assumed spectral indices and at a frequency
of 8 GHz, which is an order of magnitude higher than the typical
observed redshifted 21-cm frequency.}, and overestimates could explain
the five 21-cm absorbing galaxies exhibiting a large core
prominence. Nevertheless, from their CSS and GPS sample \citep{gs06a}
there is a 21-cm detection rate of 1 out of 9 for quasars, cf. 15 out
of 23 for galaxies, supporting the existence of a bias. Quantifying this, the binomial
probability of 15 or more out of 23 detections occurring in one
class, while 8 or more non-detections occur in another class is
0.20\%, again consistent with quasars being the result of lower
inclined obscuring tori.

\subsubsection{Quasar--galaxy classifications}
\label{sec:class}

The classification schemes used by these authors do not apply to all the sources under
consideration in this work. We have therefore applied our own scheme
in order to examine the effect of the relative strength of the
quasar/AGN nucleus over the host galaxy. The aim is to distinguish
sources whose appearance is dominated by the nuclear source (the
``quasars'') from those whose appearance is dominated by the extended
stellar light of the host galaxy (the ``galaxies''). The latter
sources are those with either intrinsically weaker AGN or the type-2
AGN that are obscured from the line-of-sight. The classification was
mostly done on a morphological basis, using SuperCOSMOS Sky Survey
images and classification (which does a star/galaxy classification
based on the source profile), DSS2 images, as well as detailed imaging
and spectroscopic analysis (where the spectra are examined for
evidence of galaxy light or quasar emission lines) available from the
literature. Each source was examined individually to ensure the
classifications were consistent.
There are some obvious caveats with this process. By looking at the
morphology, we are biasing ourselves somewhat against high-redshift
galaxies, since we lose the resolution and surface-brightness
sensitivity for detecting the galaxy light. However, many of the
high-redshift sources have higher-resolution imaging or spectroscopic
observations available that allow us to determine the importance of
the host galaxy in the source's appearance.

In order to compare the proportions of detections and non-detections for  
galaxies and quasars, we use the following statistic.
If we have two measured proportions $\hat{p}_1=X_1/N_1$ and $\hat{p}
_2=X_2/N_2$, with the total proportion
$\hat{p}=(X_1+X_2)/(N_1+N_2)$, then
\begin{equation}
\label{eq:1}
T = \frac{\hat{p}_1 - \hat{p}_2}{[\hat{p}(1-\hat{p})(N_1^{-1}+N_2^{-1})]^{1/2}},
\end{equation}
which has a standard normal distribution (mean zero, variance one) under  
the null hypothesis, being that the two proportions ($\hat{p}_1$ and $\hat{p}_2$) are the  
same. We examined the proportion of
sources that we classified as galaxies (as opposed to quasars), and
found that sources with detected absorption had a much higher galaxy
proportion (28 out of 34) than sources without absorption (28 out of
56). These fractions are different at 99.67\%
confidence. Additionally, almost all the high-$L_{\rm UV}$ sources
are ``quasars'' -- only 1 out of the 24 sources with
$L_{\rm UV}>10^{22.5}$ \WpHz\ is a galaxy.

\subsubsection{Our results in the context of the unified schemes}

All of the above discussion on orientation effects applies to the low
redshift sample, where searches have previously been concentrated
(Fig.~\ref{tau-z})\footnote{The large gap in redshift space between
  these and the high redshift targets (Fig.~\ref{tau-z}) is due to a
  lack of (RFI free) coverage over $\approx350$ to 700 MHz.}.  Of our
own low redshift searches (Paper I), one target out of three was
detected in 21-cm absorption, i.e.  as per the 33\% detection rate of
\citet{vpt+03}, the rate expected from unified schemes\footnote{The
  detection rate for a given half-opening angle, $\alpha$, is
  $1-\cos\alpha$, giving a detection rate of 29\% for a randomly
  oriented population with a $90^\circ$ opening angle.}. For the two
undetected, 1450$-$338 ($z=0.368$) has an estimated UV luminosity of
$L_{\rm UV}=4\times10^{18}$ W Hz$^{-1}$ and 2300$-$189 ($z=0.129$)
$L_{\rm UV}=1\times10^{20}$ W Hz$^{-1}$, cf. $L_{\rm
  UV}=5\times10^{19}$ W Hz$^{-1}$ for the detection in 1555$-$140
($z=0.097$). That is, the UV luminosity of the detection lies between
those of the non-detections, although all lie well within the $L_{\rm
  UV}\lapp10^{23}$ W Hz$^{-1}$ detection range (Fig.~\ref{ion}).
Regarding their orientations, 1450$-$338 is an apparently
dust-reddened quasar \citep{fww00}, 2300$-$189 is a Seyfert 1 galaxy,
whereas the strong ($N_{\rm HI}=4.2\times10^{19}.\,T_{\rm s}/f$ \scm)
21-cm absorber 1555$-$140 has a reddened type-2 AGN spectrum
superimposed on a galaxy \citep{wwjp83}, and is itself a large galaxy
in the centre of a group. With the one detection occurring in the only
type-2 object, our low luminosity results are consistent with unified
schemes, as discussed above.

However, although our high redshift criterion selects highly luminous
sources, note that half (8 out of 16) of those with $L_{\rm
  UV}\gapp10^{23}$ W Hz$^{-1}$ have redshifts of $z_{\rm
  em}\leq0.73$\footnote{This demonstrates that the segregation of our
  high luminosity sample is not due to a systematic underestimate in
  the low redshift UV luminosities, where the extrapolation of the
  SEDs are generally more extreme than for the high redshift sources
  (Sect. \ref{uvf}).} and, in common with the other $L_{\rm
  UV}\gapp10^{23}$ W Hz$^{-1}$ targets, these are all non-detections.
This raises the question of whether orientation effects alone can
account for the observed differences. That is, why do some of the low
redshift non-detections exhibit low UV luminosities when, like their
high luminosity counterparts, we expect to have a direct view to the
UV emitting region in these type-1 objects?

\subsection{The significance of known intervening absorbers}
\label{kia}

Most of our targets also have intervening DLAs and so if the photons
emitted from the quasar are ionising, a significant number of these
must have been redshifted to 1216 \AA\ by time they encounter the
absorber, providing a continuum for the Lyman-\AL\ absorption. In fact,
DLAs could be common to many of the objects searched for host
absorption, although ground based observations of the Lyman-\AL\ line
are usually restricted to redshifts of $z\geq1.7$, where the line is
redshifted into optical bands\footnote{Additionally, at redshifts of
$z\gapp4$, the increasing line density per unit redshift of the
Lyman-\AL\ forest, makes the identification of DLAs very difficult.}.
From Fig. \ref{tau-z}, we see that only our targets, the two high-$z$
21-cm detections (\citealt{ubc91,mcm98}, Table \ref{t1}) and the two
targets of \citet{cwh+07} are of sufficiently high redshift to
illuminate such absorbers. The DLAs which intervene our targets (nine
in total towards seven sources) are all themselves at redshifts
($z_{\rm abs}\geq1.947$), comparable to those of the background
quasars (Table \ref{dlat}).
\begin{table}
\centering
\begin{minipage}{72mm}
\caption{The targets with detected intervening DLAs ($N_{\rm
HI}\ge2\times10^{20}$ \scm) and sub-DLAs. Note that 0913+003,
1026--084 \& 1251--407 have not featured in any of our prior
statistics since the \HI\ search in 0913+003 was ruined by RFI and in
the latter two we searched for OH only (\HI\ is redshifted out
of the band).
\label{dlat} }.
\begin{tabular}{@{}l c c c c c  }
\hline
QSO & $z_{\rm abs}$ & $\log N_{\rm HI}$ & Ref. & $z_{\rm em}$ & $\Delta z$ \\
\hline
0335--122 & 3.178 & 20.8 & E01 & 3.442 & 0.063 \\
0347--211 & 1.947 & 20.3 & E01 & 2.944 & 0.338  \\
0537--286 & 2.974 & 20.3 & E01 & 3.104 & 0.033 \\
0913+003  & 2.774 & 20.3 & E01 & 3.074 & 0.079 \\
1026--084 & 3.42  & 20.1 & P01 & 4.276 & 0.193 \\ 
...	  & 4.05  & 19.7 & P01 & ...   & 0.045 \\
1228--113 & 2.193 & 20.6 & E01 & 3.528 & 0.418 \\
1251--407 & 3.533 & 20.6 & E01 & 4.464 & 0.205 \\
..        & 3.752 & 20.3 & E01 & ...   & 0.150\\
\hline
\end{tabular}
{References: E01: \citet{eyh+01}, P01: \citet{psm+01}.\\ Notes: No DLAs
have been found towards 1351--018, 1535+004, 1937--101, 2215+02, 2245--059
\citep{eyh+01}, 1937--101 \citep{lwt+91}.}
\end{minipage}
\end{table}

The redshift of the background source in the rest frame of the
absorber is given by $\Delta z = \frac{z_{\rm em}+1}{z_{\rm abs}+1} -
1$ and at $\Delta z=0.33$, the 912 \AA\ (ionising) photon is
redshifted to 1216 \AA. All but two of the DLAs have $\Delta z<0.33$
(Table \ref{dlat}), meaning that the radiation which is
Lyman-\AL\ absorbed must have been non-ionising at the source.  Two of
the DLAs have also been searched, and not detected, in 21-cm
absorption; 0335--122 and 0537--286 \citep{kc02}. The non-detections
place spin temperate/covering factor ratios of $T_{\rm s}/f\gapp2000$
and $\gapp700$ K, respectively, and since both occult very compact
radio sources (Table 2 of \citealt{cmp+03}), $f$ may be close to
unity, indicating high spin temperatures in these absorbers
(cf. \citealt{cw06}). Therefore, as shown by these two cases, the
non-detection of 21-cm does not necessarily imply a lack of neutral
gas close to the quasar.

In the absence of total neutral hydrogen column densities, we cannot
estimate the limits for our targets (Table \ref{sum}), although for
$N_{\rm HI}\gapp10^{20}$ \scm, $T_{\rm s}\gapp10^3$~K, which is around
the typical upper limit for the detection of 21-cm absorption in DLAs
(figure 5 of \citealt{ctp+07}).  The spin temperature measures the
relative populations of the two possible spin states \citep{pf56},
although excitation to the $n=2$ level by Lyman-\AL\ photons will also raise
the spin temperature \citep{fie59}.  Furthermore, \citet{be69} show
that both the 21-cm (cf.  Fig. \ref{lum}) and the Lyman-\AL\ (cf.
Fig. \ref{ion}) flux can contribute to the spin temperature at
absorber--quasar separations of less than a few tens of kpc, i.e. for
associated systems.  

In addition to many of the background quasars of our sample emitting a
large fraction of non-ionising, high energy ($1216 < \lambda < 912$
\AA) photons, there exists a number of proximate damped Lyman-$\alpha$
absorption systems (PDLAs), where at $\Delta v\leq3000$ \kms, the
ionisation of the gas is expected to be dominated by the quasar,
rather than the background UV flux\footnote{At $10^4\gapp\Delta
v\gapp10^5$ \kms\ (Table \ref{dlat}), none of the DLAs intervening our
targets are proximate.}. Unlike the detections in the associated
sample (Table \ref{t1}), many of the PDLAs are subject to $L_{\rm
UV}\gapp10^{23}$ W Hz$^{-1}$ (Fig. \ref{pdla-lum}).
\begin{figure}
\vspace{6.8cm}
\includegraphics{pdla-lum.ps}
\caption{The ultra-violet ($\approx912$ \AA) luminosity versus the
velocity offset of the Lyman-\AL\ absorption from the background
quasar for the SDSS DR5 PDLA sample \citep{phh07}. The quasar
redshifts range from $z_{\rm em} = 2.308$ to $5.185$ and the
luminosities have been estimated as per the associated sample
(Sect. \ref{uvf}). The solid symbols represent blueshifted
(approaching) \HI\ and the hollow symbols redshifted (receding) \HI.  The
horizontal lines signifies the detection cut-off for the associated
systems and the vertical line, the fiducial $300$ \kms\ offset above
which the gas is believed to be unassociated.}
\label{pdla-lum}
\end{figure}
From the figure we see that few of the PDLAs appear to arise in the
host (i.e. at $\Delta v\lapp300$ \kms). Comparing these with the
associated sample (Fig.~\ref{tau-vel}), a Kolmogorov-Smirnov test on
the velocity offsets gives a probability of only $9.7\times10^{-12}$
that the associated detections and the PDLAs are drawn from the same
sample. Therefore, although there are several PDLAs for which $L_{\rm
UV}\gapp10^{23}$ W Hz$^{-1}$ at $\Delta v\lapp300$ \kms, PDLAs are not
(generally, at least) associated systems (see also \citealt{mwf98}). 
That is, although it
is feasible that a search for associated absorbers could overlap
significantly with a search for PDLAs, we show that this is generally not the case.
For the PLDAs, $\Delta z\lapp0.01$ and, as per the DLAs detected towards our targets
(Table \ref{dlat}), this would suggest again that much of the flux
from the quasar is non-ionising at the source\footnote{Since
DLAs do not generally occult quasars of sufficient radio flux to be
detected in 21-cm absorption at present ($\gapp0.1$ Jy, see
\citealt{cwbc01}), it is likely to remain unknown for some time
whether these PDLAs are host to {\em cold} neutral gas.}.

Since, by definition, the DLAs towards our targets are subject to a
high 1216 \AA\ flux, there must be a significant portion of ionising
photons present, which could render the gas less detectable in
Lyman-\AL\ (and 21-cm) absorption, although at these high column
densities self-shielding effects (\citealt{zm02}) should counteract
this somewhat.  Note, however, that none of the sight-lines towards our targets
exhibit further DLAs between those listed in Table~\ref{dlat} and
$z_{\rm em}$, where the ionising photons will be less redshifted. In
Fig. \ref{N-uv} we convert the luminosities to fluxes for the DLAs
towards our targets and the PDLA sample.
\begin{figure}
\vspace{6.8cm}
\includegraphics{N-uv-all.ps}
\caption{The neutral hydrogen column density versus the incident UV
  flux at the DLAs towards our targets (Table \ref{dlat}, unfilled
  markers) and the $\Delta v\geq300$ \kms, $z_{\rm abs}<z_{\rm em}$
  PDLAs (\citealt{phh07}, filled markers). These criteria have been
  applied in order to select absorbers directly along the
  line-of-sight and with the redshifts not dominated by peculiar
  motions. The scatter in the distribution remains if the whole
  PDLA sample is used.}
\label{N-uv}
\end{figure}
Although, there is a tentative trend for the column density of the
neutral gas to decrease with flux for our targets, this is not borne
out by the larger PDLA sample. Although this would suffer a selection
effect by being limited to the $N_{\rm HI}\ge2\times10^{20}$ \scm\ DLA
defined cut-off (where self-shielding becomes particularly
significant, \citealt{zm02}), at higher incident fluxes there appears
to be no additional appreciable photo-ionisation. However, PDLAs are
considerably less numerous than expected, which \citet{phh07}
attribute to photo-evaporation. Another factor which could cover any
neutral gas column density--incident UV flux anti-correlation is the
possibility that the observed redshift differences between the
absorbers and quasars do not provide accurate distances in these
cluster environments. This could be the case for the $\Delta v\leq300$
\kms\ PDLAs of Fig. \ref{pdla-lum}.

\subsection{The chicken or the egg: ionising UV flux or orientation effects}

\subsubsection{Orientation: Ionisation}

Since Lyman-$\alpha$ absorption is not detected in the hosts of any of our
high redshift quasar sample \citep{eyh+01,psm+01}, it is not surprising that 21-cm
absorption remains undetected.  Naturally, intervening DLAs require
sufficient background UV flux against which to detect absorption, and
by selecting such high redshift sources, we are selecting those which
are sufficiently luminous in the UV to enable the detection of
intervening absorption, but perhaps too bright to host a large column
of neutral gas close to the host.

This raises the question of whether it is the UV luminosity ionising
the neutral gas, which would otherwise be there, or the
distribution of the gas (a face-on torus), which is responsible for
the non-detection of 21-cm absorption in these sources. For the PDLAs
there are several cases where $L_{\rm UV}\gapp10^{23}$ W Hz$^{-1}$ in
what could be associated absorption ($\Delta v\leq300$ \kms) and
orientation effects could be consistent with the complete lack of any
PDLAs in our own (admittedly, small) target sample (Table
\ref{dlat}). This paucity may however be expected without invoking
orientation effects, since PDLAs constitute only a small fraction
($\sim5\%$) of the DLA sample \citep{eyh+02,phh07} and those which
may be associated ($\Delta v\leq300$ \kms) are even rarer
($\lapp1\%$).

\begin{figure*}
\centering
\includegraphics[angle=270,scale=0.70]{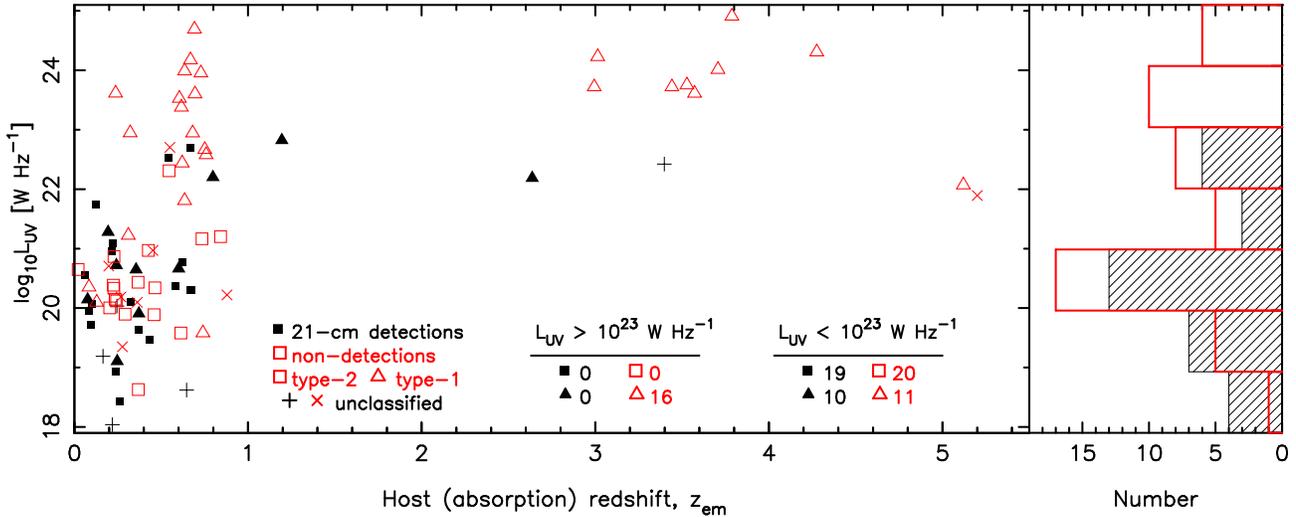}
\caption{The ultra-violet luminosity--redshift distribution for the
  sample. The symbols and histograms are as per Fig. \ref{lum}, but now the shapes
  represent the AGN classifications, with triangles representing
  type-1 objects and squares type-2s -- the legend shows the number of
  each according to the UV luminosity cut-off (cf. Table
  \ref{totals}). The distribution indicates that the extrapolations
  used for the low redshift sources (Sect. \ref{uvf}) do not appear to
  introduce an overwhelming bias in comparison to the high redshift
  sources (at least for the quasars).  Should the extrapolations be
  overly erroneous, this would result in a systematic shift in the
  luminosities of the low redshift sample, although there would still
  be the sub-sample of those undetected in 21-cm absorption at higher
  UV luminosities.}
\label{lum-z}
\end{figure*}

In AGN, large-scale (kpc) outflows of ionised gas, directed along the
radio jets are rife (particularly in Seyfert galaxies, see table 1.2
of
\citealt{cur99}\footnote{http://nedwww.ipac.caltech.edu/level5/Curran/frames.html}).
These are believed to be due to either nuclear gas which is ionised
and driven out along the radio jets
(e.g. \citealt{bbr84,sch88,cbg+96,cbov98}) or photoionised ambient
galactic gas \citep{pdu85,upa+87,fws98}. This latter
model\footnote{Where the torus may be a consequence of the weak
radiation emitting from the equator of the continuum source, with the
cone arising from gas ionised by the strong polar radiation
\citep{pfh+98}.} is consistent with the UV radiation ionising all of
the gas, while giving a high UV luminosity in our
direction. Furthermore, the paucity of extended ionised structures in
type-1 Seyfert galaxies \citep{pog89b}, which unified schemes dictate
are intrinsically identical to their type-2 counterparts, suggests
that these are being viewed at low inclinations and so do not extend
beyond the central AGN from our viewpoint.  Furthermore, {\em at all
redshifts} (Fig. \ref{lum-z}), all of the $\log L_{\rm UV}\gapp23$
radio sources have been flagged as quasars (by us as well as by
\citealt{vpt+03,gss+06})\footnote{Although having classified all of
the sources for ourselves (Sect. \ref{sec:class}), not all of the assigned
designations agree.}. That is, the sources in which the UV flux
appears to be directed towards us are believed to be type-1 objects,
in addition to having a very low (possibly zero) likelihood of
exhibiting 21-cm absorption.


Although random orientation effects could give the 50\% split of 21-cm
detections and non-detections below the median value of $\log L_{\rm
UV}< 21.5$ (Sect. \ref{uvf}), at higher UV luminosities the number of
21-cm detections drops significantly, especially at $\log L_{\rm
UV}\gapp23$.
\begin{table} 
\caption{Partitioning of the 21-cm detections and non-detections with
respect to $L_{\rm UV}=10^{23}$ W Hz$^{-1}$.}
\begin{tabular}{@{}r r c c c c|c c c@{}} 
\hline
 & \multicolumn{3}{|c|}{$L_{\rm UV}\lapp10^{23}$ W Hz$^{-1}$} & \multicolumn{3}{|c|}{$L_{\rm UV}\gapp10^{23}$ W Hz$^{-1}$} &\\
  & dets & nons & total & dets & nons & total \\
\hline
Galaxies & 27 & 24 & 51 &   --  & -- &  0\\
Quasars  & 6  & 12 & 18 &   0   & 16 &  16\\
Total    & 33 & 36 & 69 &   0   & 16 & 16 \\
\hline       
\end{tabular}
\label{totals}  
\end{table} 
%
Also, for the low luminosity sample, there is a detection rate of 1/2
for the galaxies and 1/3 for the quasars (Table~\ref{totals}) and,
while this is consistent with previous studies \citep{pcv03}, the fact
that the detection rate for the low UV luminosity quasars is
significantly greater than for the $L_{\rm UV}\gapp10^{23}$ W
Hz$^{-1}$ quasars ($0/16$), suggests that these may be different a
beast than their low UV luminosity counterparts.


%
%
This is also illustrated in Fig. \ref{lum-z}\footnote{The symbols are
  discussed in the next section.}, where it is seen that all of
$z_{\rm em}\gapp1$ targets have UV luminosities in excess of $L_{\rm
  UV}\sim10^{22}$ W Hz$^{-1}$, demonstrating that there is a selection
effect at play, where at high redshifts we are targetting the most
optically bright sources, which are naturally better known at such
high luminosity distances. At $z_{\rm em}\lapp1$ (luminosity distances
$\lapp6$ Gpc), we see the whole range of luminosities. Again, the fact
that all of the detections occur at $L_{\rm UV}\lapp10^{23}$ W
Hz$^{-1}$ is clearly evident, but whether the difference in
luminosities is an intrinsic property is difficult to ascertain: The
low UV luminosity quasars undetected in 21-cm absorption may be at
sufficiently low inclinations so that a large column of neutral gas is
not intercepted by the radio emission, while being inclined highly
enough so that the axis of the jet does not cross our line of sight,
i.e. the UV radiation is not directed towards us in these
cases. Considering this, these low luminosity quasars may represent
intermediate types (1 to 1.5, \citealt{kee80,mr95b}), which are known
to have stronger ionisation lines than type-2 objects \citep{coh83}.

\subsubsection{Orientation: Obscuration}
\label{o:o}

If the orientation is important in the detection rate of \HI\
absorption, one manifestation of it should be a higher extinction of
the quasar emission by dust associated with the torus. We can test
this via an optical--near-infrared colour-colour diagram, using the
photometry in  Appendix~A,
\begin{figure}
\vspace{6.7cm}
\begin{picture}(0,0)
\includegraphics{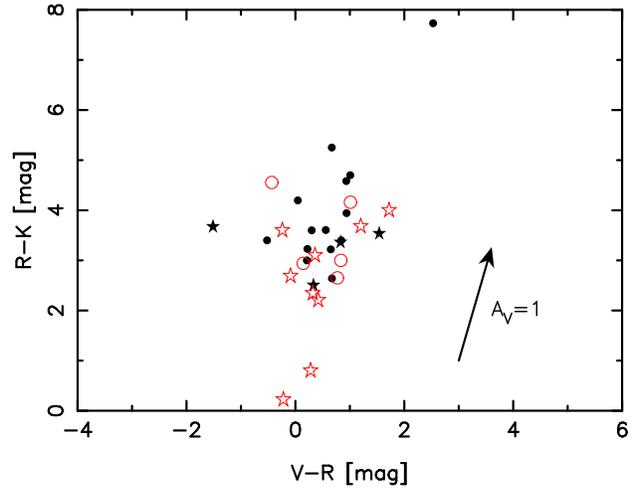}
\end{picture}
\caption{The $R-K$ colour versus the $V-R$ colour for the sample. As per
Fig. \ref{lum}, the filled symbols represent the 21-cm detections and
the unfilled symbols the non-detections.}
\label{colourcolour}
\end{figure}
 and in Fig. \ref{colourcolour} we show the $V-R$ and $R-K$ colours, where
available, for the sample.  From this there is little
apparent difference between the two populations: Although there are a
few sources which lie off the main distribution in the direction
expected from dust reddening, the bulk of the detections and
non-detections are found to be concentrated in the same part of the
plot. This indicates that the detections are not significantly more
dust reddened than the non-detections. While this might indicate that
the orientation hypothesis is not supported, it should be noted that
if the dusty torus did significantly extinguish the quasar light, 
the host galaxy starlight would then dominate, lessening
the apparent reddening. 

As a check on the effect of orientation, we have examined the spectral
type for each source in the full sample. This entailed an exhaustive
literature search to find, ideally, published measurements of emission
line fluxes, or alternatively, a published spectrum (Tables
\ref{tab:photDet} and \ref{tab:photNon}). This information was then
used to classify the AGN as a type-1 (i.e. showing broad permitted
lines), or a type-2, where only narrow lines are present.  The result
is shown in Fig. \ref{lum-z}, where the detection rate for type-1
objects is 10/37, compared to 19/39 for the type-2s, indicating a
preference for type-2 objects to exhibit 21-cm absorption.

If we split the sample at $L_{\rm UV}=10^{23}$~\WpHz, we see
that all 16 of the high-luminosity sources are found to be type-1
objects.  Being exclusive non-detections, this is in line with unified
schemes.  However, for the low luminosity sources, the detection rates
are close to 50\% for both types (10/21 for type-1s and 19/37 for
type-2s).  So while the raw detection rates for type-1s and type-2s
appear different (actually only at a $1.95\sigma$ level), the dominant
cause of the different detection rates between the type-1 and type-2
objects appears to be due to the 16 high-luminosity type-1 objects.
Without these, i.e. for $L_{\rm UV}<10^{23}$~\WpHz, both AGN types
have a $\sim50\%$ probability of exhibiting \HI\ absorption. In other
words, a type-1 object does not automatically result in a non-detection of
\HI\ absorption, nor does a type-2 necessarily result in a detection. This flies in the
face of the notion that the strength of the 21-cm absorption is determined by the
aspect of the central obscuring torus (Sect. \ref{oeff}) and is strong
evidence that the absorption occurs beyond the parsec-scale, possibly
in the main disk of the galaxy, the orientation of which appears to
have little bearing on the AGN type.

\subsubsection{Host galaxy}

The third point to consider is thus the quasar host galaxy. The
detection of associated absorption simply requires that there is neutral
gas at similar velocities to the quasar nucleus. While we have been
discussing the prospect of gas associated with the active nucleus (for
instance, with the surrounding torus) and whether the
nuclear obscuration plays a part, the r\^{o}le of gas in the host galaxy
needs to be considered. Quasars, particularly at high redshifts, are
observed to reside in galaxies of diverse types: For example, \citet{pir+06}
find that quasar hosts at $1<z<4.5$ span a range of morphologies
consistent with early-types to disky/late-type galaxies. While some early-type
galaxies are known from targeted searches to have significant
\HI\ content, particularly in the field
(e.g. \citealt{mdo+06,omd+07}), blind surveys, such as the ALFALFA
survey \citep{dgg+07}, show that early type galaxies in clusters have
a much lower neutral gas content. Those quasars with detected \HI\
absorption are then more likely to be found in disky or
late-type galaxies.

Our observed trend in the detection rate as a function of UV
luminosity could thus be explained by a changing mix of host galaxy
types, in the sense that a larger fraction of the more luminous
quasars are found in early-type galaxies. While observational
constraints limit our knowledge of such a tendency at $z>2$, this
trend is known to apply at lower redshifts. For instance,
\citet{tdhr96} find that the hosts of all radio-loud quasars studied,
as well as the most powerful radio-quiet quasars (all with $z<0.35$),
have a de~Vaucouleurs $r^{1/4}$ law profile (characteristic of
elliptical galaxies), whereas the less powerful radio-quiet quasars
exhibited exponential disc profiles. \citet{tdhr96} suggest that the
most luminous quasars reside in elliptical galaxies, regardless of
their radio properties.  More recent observations of
$z<0.3$ quasars, using adaptive optics \citep{gss06}, find that most luminous
quasars (with $L>2L^*_H$) have elliptical host galaxies (this includes
the majority of the radio-loud quasars). To summarise, in this
scenario the most luminous quasars of our sample are in early-type
galaxies, which have a lower neutral gas content than later types, and
so the chance of detecting significant \HI\ absorption is greatly
reduced compared to the less-luminous quasars.

\section{Possible effects in the non-detection of molecular absorption}
\label{nomol}

In light of the absence of atomic absorption due to the large UV
luminosities, possibly due to unfavourable orientations, it is not
surprising that molecular absorption remains undetected. However,
were \HI\ detected, a detectable molecular abundance would still not
necessarily be expected from this sample, since:
\begin{enumerate}
	\item The relationship between molecular fraction and
          optical--near-infrared colour (Paper I) suggests that our sources would simply
          not be red enough to be detectable in molecular absorption
          at these sensitivities. The current sample had been selected
          before the conclusions of Paper I had been fully formulated, and
          so the range of $V-K$ colours most likely lie off to the left
          in Fig.~\ref{OHoverH} (the colours range from $V-K=1.08 - 2.63$,
          where available).
\begin{figure}
\vspace{8.0cm} \setlength{\unitlength}{1in} 
\begin{picture}(0,0)
\put(-0.2,3.85){\includegraphics{OHoverH-all-V.ps}}
\end{picture}
\caption{The normalised OH line strength $\left(2.38\times10^{14}\int
\tau_{_{\rm OH}}\, dv/1.82\times10^{18}\int\tau_{_{\rm HI}}\,
dv\right)$ versus optical--near-IR colour for the known OH absorbers.
These are represented by the filled symbols, with the least-squares
fit shown for the four millimetre systems. The correlation for the
five OH absorbers is significant at the $2.0\sigma$ level, which rises
to $3.0\sigma$ for the molecular fraction/$V-K$ correlation for these
sources plus the \MOLH-bearing DLAs (figure 1 of \citealt{cwm+06}).
The unfilled symbols show the other sources where \HI\ has been
detected and OH absorption searched: The three limits shown in figure
7 of \citet{cwm+06} [green] (the lower limits on the abscissa
designate $R-K$ magnitudes) with a further three from
\citet{gss+06} [blue]. Note that the OH limits have been rescaled
according to the method described in \citet{cdbw07}, who find ${\rm
FWHM}_{\rm OH}= {\rm FWHM}_{\rm HI}$ for the five known OH absorbers:
Combining the otherwise unknown ${\rm FWHM}_{\rm OH}$ with the optical
depth limit, gives a more accurate estimate for the upper limit than
quoting the column density limit per each $\Delta v$ channel. This has
the effect of degrading the apparent sensitivity, although some can be
recovered since usually ${\rm FWHM}_{\rm OH} > \Delta v$. Therefore
for the non-detections, we scale each of the OH column density limits
by $\sqrt{{\rm FWHM}_{\rm OH}/\Delta v}$, thus giving the limit for a
single channel ``smoothed'' to FWHM$_{\rm OH}$. Associated absorption
is designated by a square and intervening absorption (due to a gravitational
lens) by a triangle.}
\label{OHoverH}
\end{figure}
	\item Furthermore, due to the metallicity and molecular
	fraction evolution noted in DLAs \citep{cwmc03}, at $z\geq3$ we may
	expect much lower abundances than present day values.  From a
	search of millimetre lines in DLAs with the Green Bank
	Telescope, \citet{cmpw03} reached similar limits as this survey
	(Table~\ref{tid-sum}). However, upon applying the molecular
	fraction evolution to the limits, they found that
	the survey was only sensitive of molecular fractions of close
	to unity, a value which even the DLAs detected in \MOLH\ fall
	very far short of, although for $V-K\gapp5.3$ such high
	fractions may be expected (figure 1 of \citealt{cwm+06}).

	\item Lastly, at such high redshifts the cosmic microwave
          background will raise excitation temperatures
	of $T_{\rm x}=10$ K at $z=0$ to $\gapp20$ K. This has the
	effect of decreasing the sensitivities to these ground state
	transitions -- by a factor of two for OH 18-cm ($^{2}\Pi_{3/2}
	J = 3/2$) and by up to a factor of four for the
	$J=0\rightarrow1$ transitions \citep{cmpw03}. One solution to
	this is to search higher transitions redshifted into the 3-mm
	band (cf. the 12-mm band), although lower flux densities,
	compounded with the need for much better observing conditions,
	makes this a poor trade.
\end{enumerate}

\section{Summary}

We have undertaken a survey for \HI\ 21-cm and rotational molecular
absorption in the hosts of radio sources at redshifts of $z\geq2.9$,
and report no detections in the 13 sources for which we have good
data. Upon comparing our search criteria with those formulated in
Paper I \citep{cwm+06}, we are not surprised that molecules (OH, HCN,
HCO$^+$ \& CO) were not detected in the 10 separate sources searched:
With optical--near-infrared colours of $V-K\leq2.63$ (at least where
these are available), the sources do not exhibit the degree of
reddening indicative of the dust abundances which would permit
molecular fractions of close to unity, the limit to which current
radio observations are sensitive.

However, since the co-moving density of \HI\ at $z \sim 3$ is expected
to be many times higher than present values (e.g. \citealt{psm+01}),
the absence of detections of 21-cm absorption was surprising.  We rule
out the possibility that the non-detections are due exclusively to
high 1420 MHz continuum fluxes maintaining an overpopulated 21-cm
upper hyperfine (anti-parallel) level. We do find, however, that all of our
targets are quasars and the ultra-violet continuum luminosities are in
excess of $L_{\rm UV}\sim10^{23}$ W Hz$^{-1}$. In comparison to the
previous searches for redshifted 21-cm absorption:
\begin{itemize}
  \item A mix of 21-cm detections and non-detections at lower
    redshifts (mostly $z\lapp1$) is well documented, where the
    detection rate is higher in galaxies than in quasars. This skew in
    the distribution is attributed to the possibility that radio
    galaxies are type-2 sources, whereas quasars are type-1
    objects. That is, the more direct view to the active nucleus in a
    quasar means that the dusty obscuring torus, invoked by unified
    models, is orientated so that 21-cm absorption does not occur
    along our sight-line. Although our own low redshift results (Paper
    I) are consistent with this scenario, we find little evidence for
    a significantly higher degree of dust reddening by any obscuring
    gas in the 21-cm detections.
    
    \item Despite the fact that half of the whole $L_{\rm
      UV}\gapp10^{23}$ W Hz$^{-1}$ sample are at $z\leq0.73$, this is
      the first time that a UV luminosity bias has been noted. The
      high UV luminosity again may be consistent with these being
      type-1 objects, where the bright UV continuum suggests we are
      seeing the accretion disk directly, unobscured by the
      circumnuclear torus.  However, although random orientation
      effects can explain the mix of detections and non-detections in
      the low luminosity sample, the high luminosity sample, in which
      there are no detections, remains unexplained. This suggests that
      there are additional effects at play.
  \end{itemize}

As well as several of our high redshift targets having intervening
DLAs at similar redshifts to the background quasar, there exists a
sample of proximate DLAs, where $z_{\rm abs}\sim z_{\rm em}$, and both
of these groups show that large columns of neutral gas can in fact
exist close to $L_{\rm UV}\gapp10^{23}$ W Hz$^{-1}$ QSOs. Since the
gas in these intervening absorbers is not associated, and are thus
free to have any aspect with respect to the AGN, this does not
contradict the possibility that our non-detections are the result of
orientation between the associated gas and the line-of-sight to the
quasar. Furthermore, no other absorbers have been found closer to our
targets \citep{eyh+01,psm+01}, perhaps suggesting that either this gas
is unfavourably orientated or that there is a proximity effect, where
the intense UV radiation is photo-ionising the associated gas clouds
\citep{bdo88,phh07}. This latter scenario could also be responsible
for high spin temperatures in the DLAs towards our targets, which are
sufficiently remote to host large columns of neutral gas, two of
which have been searched and not detected in 21-cm absorption
\citep{kc02}.

Finally, although 21-cm absorption shows a slight preference to be
present in galaxies over quasars, by determining the AGN
spectroscopic type for each object (a total of 76), we find that below
$L_{\rm UV}\sim10^{23}$ W Hz$^{-1}$ {\em the presence of 21-cm absorption
  shows no preference for AGN type}.That is, both type-1 and type-2
objects have a 50\% likelihood of exhibiting 21-cm absorption and
any apparent bias against type-1 objects is due solely to the 16 
$L_{\rm UV}\gapp10^{23}$ W Hz$^{-1}$ objects. This means:
\begin{enumerate}
  \item The ultra-violet luminosity, rather than the orientation of
    the AGN, can determine whether 21-cm absorption can be detected
    in the host galaxy, where at high luminosities 21-cm is never detected
    and at low luminosities the odds are even.
    \item The \HI\ absorption probably does not occur in 
      the obscuring torus, but in the large-scale galactic disk or is
      possibly associated with in-falling or out-flowing material
      (e.g. \citealt{jm94,pvtc99,mot+03,vpt+03}). As such, there is as yet no definite
      explanation why there is only a 50\% detection rate in $L_{\rm
        UV}\lapp10^{23}$ W Hz$^{-1}$ sources.
      \item If our classifications are to be trusted, although type-1 objects
        are more likely to arise in quasars and type-2s in galaxies, this
        is not the case for every object.
\end{enumerate}
Whether all UV luminous sources arise in type-1 objects, is difficult
to ascertain, and the fact that there are low UV luminosity
non-detections is somewhat of a puzzle. If, as we believe, the
absorption is occuring in the galactic disk, it may be the orientation
of this which is responsible for the observed 21-cm optical
depths. This would suggest that the galactic disk does not necessarily
share a similar orientation to the dense circumnuclear gas on the
parsec scale\footnote{Contradicting low redshift surveys 
(\citealt{kee80,mr95b}, and to a certain degree, \citealt{cur99p}), which
find that find that intermediate Seyferts of types 1, 1.2 and 1.5 will
occur in face-on galaxies while those of type 1.8 and 1.9 will occur
in the edge-on cases.}. Alternatively, a possible explanation for the
high luminosity non-detections is that this selection biases towards
a specific class of host galaxy, i.e. gas-poor, early types, in our
targets as well as the $L_{\rm UV}\gapp10^{23}$ W Hz$^{-1}$ quasars at
$z\leq0.73$. This may render the orientation argument also invalid in these
cases.

The exclusive non-detections of 21-cm absorption for $L_{\rm
  UV}\gapp10^{23}$ W Hz$^{-1}$ quasars indicate why we do not see
absorption in our PQFS sample. Our selection of targets was biased
towards high UV luminosity sources in several ways. Firstly, the
requirement of having a measured optical redshift preferentially
selects objects that are relatively bright in the optical
(i.e. rest-frame UV), where a suitable spectrum can be obtained in a
feasible observing time. In the PQFS, 509 of the 878 sources have
measured redshifts, so that about 42\% of the sample is unavailable
for this study. If anything, these missing sources would have
luminosities lower than the quasars of our sample and so the detection
of 21-cm absorption in any of these would have little bearing on our
result.

Secondly, our high redshift selection clearly biases towards the
brightest UV sources. This is despite our $B\gapp19$ selection, which
gives the luminosity ceiling of $L_{\rm UV}\lapp3\times10^{24}$ W
Hz$^{-1}$ at $z\sim3$\footnote{The fact that one of our targets lies
  on the flux limit with $B=19$, while also being the highest redshift
  source observed ($z=3.8$), gives the one point above $L_{\rm UV}=3
  \times 10^{24}$ W Hz$^{-1}$ (Fig. \ref{lum-z}).}, well above the
$10^{23}$ W Hz$^{-1}$ fiducial limit.
Therefore, in order to detect associated \HI\ at $z_{\rm
em}\geq3$, sources with UV luminosities of $L_{\rm UV}\lapp10^{23}$ W
Hz$^{-1}$, should therefore be targetted. At luminosity distances of
$\geq24.5$ Gpc, this corresponds to $\lambda\geq4860$ \AA\ flux
densities of $\lapp6$ $\mu$Jy, or $V\gapp22$. For the one detection of
associated \HI\ at $z_{\rm em}\geq3$, $z_{\rm em}=3.3968$ in the radio
galaxy 0902+343 \citep{ubc91}, the observed flux density is $\approx1$
$\mu$Jy, rendering this detectable towards a UV luminosity of $L_{\rm
UV}=3\times10^{22}$ W Hz$^{-1}$, which is, not surprisingly, at the
upper end of the 21-cm detections.

\section*{Acknowledgments}
We would like the thank the anonymous reviewer for their helpful and
supportive comments. Christian Henkel for the 6-cm OH Einstein
A-coefficients. Also, many thanks to Jim Lovell for performing all of
the Tidbinbilla observations and the GMRT telescope operators for
their extensive assistance.  

We acknowledge financial support from the Access to Major Research
Facilities Programme which is a component of the International Science
Linkages Programme established under the Australian Government's
innovation statement, Backing Australia's Ability.

This research has made use of the NASA/IPAC Extragalactic Database
(NED) which is operated by the Jet Propulsion Laboratory, California
Institute of Technology, under contract with the National Aeronautics
and Space Administration. This research has also made use of NASA's
Astrophysics Data System Bibliographic Services.

This work made use of the Frequently Asked Questions of the
Statistical Consulting Center for Astronomy, operated at the
Department of Statistics, Penn State University, M.G. Akritas
Director).

Funding for the Sloan Digital Sky Survey (SDSS) and SDSS-II has been
provided by the Alfred P. Sloan Foundation, the Participating
Institutions, the National Science Foundation, the U.S. Department of
Energy, the National Aeronautics and Space Administration, the
Japanese Monbukagakusho, and the Max Planck Society, and the Higher
Education Funding Council for England. The SDSS Web site is
http://www.sdss.org/.

\section*{APPENDIX A}
\label{appa}
In this section we list all of the sources used in the analysis, which
all been obtained from the literature cited in Table \ref{t1},
complete with the compiled photometry and classification.
\begin{table*}
  \centering
  \begin{minipage}{15.7cm}
    \caption{The sources detected in 21-cm absorption, listed by their B1950.0 or J2000.0 name as given in the 21-cm search paper (Table~\ref{t1}). The final columns give the estimated luminosity at
      1216\AA\ [\WpHz] and our determination of the AGN type.}
    \label{tab:photDet}
\begin{tabular}[h]{rllllllllllcll}
\hline
Source     &Class & $z_{\rm em}$   &$B$ &Ref &$V$  &Ref  &$R$ &Ref  &$K$ &Ref  &$\log L_{\rm UV}$ & Type & Ref\\
           &      &     &[mag]  &  &[mag]  &  &[mag]  &  &[mag]  &  &[\WpHz] & & \\ 
\hline
J0025-2602 &Gal &0.3220 &20.300 & 30 &---    &    &18.084 & 30 &15.674 & 61 &20.100 &2 & 75  \\
  0108+388 &Gal &0.6685 &---    &    &---    &    &22.000 & 65 &16.690 & 71 &20.309 &2 & 43  \\
J0119+3210 &Gal &0.0600 &16.271 & 30 &---    &    &14.749 & 30 &12.600 & 31 &20.548 &2 & 28  \\
J0141+1353 &Gal &0.6210 &22.327 & 30 &20.920 & 27 &20.876 & 30 &16.680 & 27 &20.777 &2 & 64  \\
J0414+0534 &Gal &2.6365 &24.100 & 29 &23.800 & 42 &21.270 &  3 &13.540 & 42 &22.188 &1 & 42  \\
J0410+7656 &Gal &0.5985 &---    &    &---    &    &21.200 & 70 &---    &    &---    &2 & 43  \\
J0431+2037 &Gal &0.2190 &22.174 & 30 &---    &    &19.085 & 30 &14.924 & 16 &18.039 &---    &     \\
  0500+019 &Gal &0.5846 &22.500 & 17 &21.350 &  8 &20.682 &  8 &15.430 & 24 &20.367 &2 & 35  \\
  0758+143 &QSO &1.1946 &19.976 & 30 &17.460 & 57 &18.972 & 30 &15.300 & 60 &22.825 &1 & 73  \\
J0834+5534 &Gal &0.2420 &18.921 & 30 &17.390 &  1 &17.180 & 46 &14.180 & 61 &20.719 &1 & 1  \\
J0901+2901 &Gal &0.1940 &19.321 & 30 &18.078 &  1 &18.600 & 12 &15.200 & 16 &21.280 &1 & 26  \\
  0902+343 &Gal &3.3980 &---    &    &23.800 &  8 &23.500 & 21 &19.900 & 21 &22.422 &---    & 20  \\
J0909+4253 &QSO &0.6700 &18.960 &  4 &19.049 &  1 &18.220 &  4 &14.860 & 60 &22.699 &2 & 1  \\
J1124+1919 &Gal &0.1650 &22.082 & 30 &21.448 &  1 &20.513 & 30 &15.930 & 16 &19.190 &---    &     \\
12032+1707 &Gal &0.2170 &18.758 & 30 &---    &    &17.327 & 30 &14.864 & 61 &20.949 &2 & 11  \\
J1206+6413 &Gal &0.3710 &21.847 & 30 &20.790 &  1 &19.910 & 55 &---    &    &19.908 &1 & 26  \\
J1326+3154 &Gal &0.3700 &21.367 & 30 &19.822 &  1 &18.882 & 30 &14.940 & 16 &19.638 &2 & 18  \\
J1347+1217 &QSO &0.1217 &16.615 & 62 &16.050 & 62 &15.718 & 30 &13.216 & 61 &21.736 &2 & 39  \\
J1357+4354 &Gal &0.6460 &---    &    &22.708 &  1 &20.951 & 30 &---    &    &18.620 &---    &     \\
J1400+6210 &Gal &0.4310 &22.137 & 30 &20.373 &  1 &19.530 & 30 &16.130 & 16 &19.459 &2 & 43  \\
J1407+2827 &Gal &0.0766 &16.345 & 30 &14.910 & 13 &14.240 & 13 &11.601 & 61 &20.144 &1 & 23  \\
  1413+135 &QSO &0.2467 &21.055 & 30 &20.000 & 33 &18.461 & 30 &14.928 & 61 &19.105 &1 & 72  \\
  1504+377 &Gal &0.6715 &---    &    &21.808 &  1 &20.800 & 69 &16.100 & 71 &20.295 &2 & 69  \\
  1555-140 &Gal &0.0970 &18.280 & 25 &16.930 & 25 &16.280 & 25 &13.060 & 25 &19.719 &2 & 79  \\
J1815+6127 &QSO &0.6010 &21.272 & 30 &---    &    &19.122 & 30 &---    &    &20.665 &1 & 76  \\
J1816+3457 &Gal &0.2448 &20.342 & 30 &---    &    &18.459 & 30 &15.525 & 61 &20.034 &---    &     \\
J1821+3942 &Gal &0.7980 &19.598 & 30 &---    &    &18.135 & 30 &15.023 & 61 &22.202 &1 & 77  \\
J1944+5448 &Gal &0.2630 &21.732 & 30 &---    &    &18.591 & 30 &15.000 & 61 &18.424 &2 & 67  \\
J1945+7055 &Gal &0.1010 &18.726 & 30 &---    &    &17.199 & 30 &13.369 & 61 &20.067 &2 & 67  \\
J2052+3635 &Gal &0.3550 &22.083 & 30 &---    &    &21.200 & 30 &---    &    &20.648 &1 & 14  \\
J2245+3941 &Gal &0.0811 &17.788 & 30 &16.550 & 58 &15.993 & 30 &12.388 & 61 &19.947 &2 & 43  \\
J2255+1313 &QSO &0.5430 &19.535 & 30 &19.590 &  1 &19.190 & 15 &---    &    &22.530 &2 & 26  \\
J2316+0405 &Gal &0.2199 &18.595 & 62 &17.440 & 62 &17.220 & 30 &13.991 & 61 &21.081 &2 & 74  \\
J2355+4950 &Gal &0.2379 &21.101 & 30 &---    &    &18.400 & 51 &15.112 & 61 &18.940 &2 & 43  \\
\hline
\end{tabular}
{\flushleft References: 
(1) SDSS DR6, \citet{aaa+08}, 
(2) \citet{awh82}, 
(3) \citet{avc+99}, 
(4) \citet{cgb+99}, 
(5) \citet{cmg+01}, 
(6) \citet{czb86}, 
(7) \citet{cgkt06}, 
(8) \citet{cb03}, 
(9) \citet{co81}, 
(10) \citet{dfs02}, 
(11) \citet{dg06}, 
(12) \citet{dbs+96}, 
(13) \citet{dl88}, 
(14) \citet{dobt00}, 
(15) \citet{dob+97}, 
(16) \citet{dob+98}, 
(17) \citet{dwf+97}, 
(18) \citet{dps+89}, 
(19) \citet{eal85}, 
(20) \citet{erp+93}, 
(21) \citet{ed92}, 
(22) \citet{ehl05}, 
(23) \citet{eh94}, 
(24) P.~Francis (priv. comm.), 
(25) \citet{fww00}, 
(26) \citet{gw94}, 
(27) \citet{ggl+04}, 
(28) \citet{gs04}, 
(29) \citet{glw+02}, 
(30) SuperCOSMOS Sky Survey, \citet{hmr+01}, 
(31) \citet{hlrv83}, 
(32) \citet{hbwm97}, 
(33) \citet{hb89}, 
(34) \citet{hjr03}, 
(35) \citet{hsj+03}, 
(36) \citet{hms78}, 
(37) \citet{ha82}, 
(38) \citet{jb91}, 
(39) \citet{kvs98}, 
(40) \citet{ksk74}, 
(41) \citet{lwt+91}, 
(42) \citet{lejt95}, 
(43) \citet{lzr+96}, 
(44) \citet{ll84}, 
(45) \citet{lnn88}, 
(46) \citet{mcbj01}, 
(47) \citet{mbis96}, 
(48) \citet{msz+03}, 
(49) \citet{msd+96}, 
(50) NASA Extragalactic Database (NED), 
(51) \citet{obm90}, 
(52) \citet{omb+06}, 
(53) \citet{pp83}, 
(54) \citet{rtn05}, 
(55) \citet{re00}, 
(56) \citet{svw65}, 
(57) \citet{san65}, 
(58) \citet{san73}, 
(59) \citet{shr+05}, 
(60) \citet{sr00}, 
(61) 2MASS, \citet{scs+06}, 
(62) \citet{sh89}, 
(63) \citet{ssb+99}, 
(64) \citet{smad85}, 
(65) \citet{sobl93}, 
(66) \citet{sfk93}, 
(67) \citet{sk93a}, 
(68) \citet{sk93b}, 
(69) \citet{sk94}, 
(70) \citet{sk96}, 
(71) \citet{srkr96}, 
(72) \citet{sww+92}, 
(73) \citet{sr01}, 
(74) \citet{tdm+02}, 
(75) \citet{tmd+93}, 
(76) \citet{vt95}, 
(77) \citet{vtrb96}, 
(78) \citet{wbg+00}, 
(79) \citet{wwjp83}, 
(80) \citet{wl78}, 
(81) \citet{wmh+02}, 
(82) \citet{zvk+97}, 
}
\end{minipage}
  \end{table*}
\begin{table*}
  \centering
  \begin{minipage}{15.7cm}
    \caption{As Table~\ref{tab:photDet} but for the sources not detected in 21-cm absorption.}
    \label{tab:photNon}
   \begin{tabular}[h]{rllllllllllcll}
\hline
Source     &Class & $z_{\rm em}$   &$B$ &Ref &$V$  &Ref  &$R$ &Ref  &$K$ &Ref  &$\log L_{\rm UV}$ & Type & Ref\\
           &      &     &[mag]  &  &[mag]  &  &[mag]  &  &[mag]  &  &[\WpHz] & & \\ 
\hline
J0003+2129 &QSO &0.4520 &21.005 & 30 &20.580 & 10 &19.650 & 10 &---    &    &20.971 &---    &     \\
  0131-001 &QSO &0.8790 &23.340 & 25 &22.500 & 25 &20.780 & 25 &16.780 & 25 &20.221 &---    &     \\
J0157-1043 &QSO &0.6160 &17.504 & 30 &---    &    &17.039 & 30 &---    &    &23.380 &1 & 48  \\
J0201-1132 &QSO &0.6690 &16.232 & 30 &---    &    &16.073 & 30 &13.860 & 37 &24.176 &1 & 75  \\
J0224+2750 &Gal &0.3102 &19.502 & 30 &---    &    &18.263 & 30 &15.250 & 44 &21.225 &1 & 23  \\
  0335-122 &QSO &3.4420 &21.018 & 30 &20.110 & 22 &20.199 & 30 &17.510 & 22 &23.722 &1 & 6  \\
  0347-211 &QSO &2.9940 &20.476 & 30 &---    &    &20.297 & 30 &17.900 &    &23.722 &1 & 35  \\
J0348+3353 &Gal &0.2430 &20.723 & 30 &---    &    &19.110 & 15 &14.390 & 16 &20.121 &2 & 23  \\
J0401+0036 &Gal &0.4260 &20.200 & 40 &19.010 & 40 &18.532 & 30 &---    &    &20.969 &2 & 9  \\
J0521+1638 &QSO &0.7590 &19.370 & 56 &18.840 & 56 &18.480 & 15 &15.380 & 60 &22.580 &1 & 26,38  \\
  0537-286 &QSO &3.0140 &19.290 & 17 &---    &    &18.789 & 30 &16.770 & 24 &24.231 &1 & 79  \\
J0542+4951 &QSO &0.5450 &18.450 & 57 &17.800 & 57 &17.210 & 15 &---    &    &22.311 &2 & 26,34  \\
J0556-0241 &Gal &0.2350 &20.968 & 30 &---    &    &19.533 & 30 &---    &    &20.150 &2 & 14  \\
J0609+4804 &Gal &0.2769 &21.198 & 30 &---    &    &18.767 & 30 &---    &    &19.349 &---    &     \\
J0655+4100 &Gal &0.0216 &15.021 & 30 &---    &    &13.996 & 30 &10.357 & 61 &20.648 &2 & 47  \\
J0709+7449 &Gal &0.2921 &19.982 & 30 &---    &    &17.540 & 19 &13.790 & 53 &19.898 &2 & 50  \\
J0741+3112 &QSO &0.6350 &16.517 & 30 &16.100 & 54 &16.322 & 30 &16.100 &  7 &23.990 &1 & 1  \\
J0815-0308 &Gal &0.1980 &18.490 & 62 &16.940 & 62 &16.797 & 30 &13.858 & 61 &20.707 &---    &     \\
J0840+1312 &QSO &0.6808 &18.370 & 80 &17.940 & 80 &17.622 & 30 &15.280 & 60 &22.947 &1 & 49  \\
J0913+5919 &QSO &5.1200 &---    &    &23.281 &  1 &24.948 &  1 &---    &    &22.071 &1 & 1  \\
J0924-2201 &Gal &5.2000 &---    &    &---    &    &25.850 & 52 &---    &    &21.893 &---    &     \\
J0927+3902 &QSO &0.6948 &17.064 & 30 &---    &    &16.486 & 30 &---    &    &23.603 &1 & 1  \\
J0939+8315 &Gal &0.6850 &---    &    &---    &    &20.140 & 44 &---    &    &---    &2 & 64  \\
J0943-0819 &Gal &0.2280 &19.401 & 30 &---    &    &18.100 & 70 &14.750 & 16 &20.868 &2 & 14  \\
J0954+7435 &Gal &0.6950 &---    &    &---    &    &21.700 & 70 &---    &    &---    &---    &     \\
  1026+084 &QSO &4.2760 &21.070 & 30 &---    &    &19.154 & 30 &---    &    &24.308 &1 & 82  \\
J1035+5628 &Gal &0.4590 &---    &    &21.244 &  1 &20.200 & 65 &---    &    &19.889 &2 & 43  \\
J1120+1420 &Gal &0.3620 &---    &    &20.935 &  1 &20.100 & 30 &17.100 & 16 &20.098 &---    &     \\
J1159+2914 &QSO &0.7290 &17.489 & 30 &18.113 &  1 &17.652 & 30 &---    &    &23.955 &1 & 78  \\
  1228-113 &QSO &3.5280 &22.010 & 17 &---    &    &19.115 & 30 &16.370 & 24 &23.754 &1 & 17  \\
J1252+5634 &QSO &0.3210 &17.760 & 56 &17.930 & 56 &17.660 & 15 &---    &    &22.949 &1 & 1  \\
J1308-0950 &Gal &0.4640 &20.767 & 30 &20.500 & 53 &18.439 & 30 &---    &    &20.340 &2 & 75  \\
J1313+5458 &QSO &0.6130 &---    &    &21.735 &  1 &20.374 & 30 &---    &    &19.581 &2 & 77  \\
  1351-018 &QSO &3.7070 &21.030 & 17 &19.696 &  1 &19.277 & 30 &17.070 & 24 &24.014 &1 & 59  \\
J1421+4144 &Gal &0.3670 &20.496 & 30 &19.330 &  1 &18.560 & 19 &15.910 & 44 &20.435 &2 & 50  \\
J1443+7707 &Gal &0.2670 &---    &    &---    &    &18.730 & 15 &---    &    &---    &2 & 26  \\
  1450-338 &Gal &0.3680 &22.520 & 25 &20.400 & 25 &19.390 & 25 &15.230 & 25 &18.629 &2 & 17  \\
J1511+0518 &Gal &0.0840 &17.993 & 30 &16.200 & 40 &16.634 & 30 &12.081 & 61 &20.353 &1 & 5  \\
  1535+004 &QSO &3.4970 &---    &    &---    &    &---    &    &19.540 & 24 &---    &---    &     \\
J1540+1447 &QSO &0.6050 &17.480 & 80 &17.000 & 80 &17.240 & 30 &13.640 &  2 &23.529 &1 & 66  \\
J1546+0026 &Gal &0.5500 &19.730 & 80 &18.900 & 80 &---    &    &16.420 & 16 &22.703 &---    &     \\
J1623+6624 &Gal &0.2030 &19.477 & 30 &---    &    &17.430 & 30 &---    &    &20.004 &2 & 63  \\
J1642+6856 &QSO &0.7510 &19.723 & 30 &---    &    &19.219 & 30 &---    &    &22.667 &1 & 43  \\
J1658+0741 &QSO &0.6210 &19.993 & 30 &---    &    &19.598 & 30 &---    &    &22.441 &1 & 79  \\
J1823+7938 &Gal &0.2240 &19.269 & 30 &---    &    &17.415 & 30 &13.866 & 61 &20.385 &2 & 32  \\
J1829+4844 &QSO &0.6920 &16.260 & 30 &---    &    &16.860 & 15 &14.250 & 60 &24.692 &1 & 26  \\
J1831+2907 &Gal &0.8420 &21.917 & 30 &---    &    &20.200 & 30 &---    &    &21.201 &2 & 68  \\
J1845+3541 &Gal &0.7640 &---    &    &---    &    &21.900 & 65 &---    &    &---    &2 & 77  \\
  1937-101 &QSO &3.7870 &18.800 & 30 &---    &    &17.188 & 30 &13.816 & 61 &24.910 &1 & 41  \\
J2022+6136 &Gal &0.2270 &19.830 & 30 &---    &    &18.146 & 30 &---    &    &20.334 &2 & 26  \\
J2137-2042 &Gal &0.6350 &20.400 & 53 &---    &    &19.286 & 30 &---    &    &21.808 &1 & 74  \\
  2149+056 &QSO &0.7400 &23.700 & 25 &22.050 & 25 &20.850 & 25 &17.170 & 25 &19.582 &1 & 68  \\
   2215+02 &QSO &3.5720 &21.840 & 25 &20.420 & 25 &20.140 & 25 &19.340 & 25 &23.613 &1 & 17  \\
J2250+1419 &QSO &0.2370 &16.760 & 80 &16.640 & 80 &17.243 & 30 &---    &    &23.616 &1 & 1  \\
  2300-189 &Gal &0.1290 &18.430 & 17 &---    &    &16.569 & 30 &13.060 & 61 &20.099 &1 & 36  \\
J2321+2346 &Gal &0.2680 &20.315 & 30 &---    &    &18.468 & 30 &14.710 & 16 &20.187 &---    &     \\
J2344+8226 &QSO &0.7350 &21.769 & 30 &---    &    &20.220 & 70 &15.850 & 16 &21.165 &2 & 43  \\
\hline
\end{tabular}
  \end{minipage}
  \end{table*}

\section*{APPENDIX B}



The ultraviolet luminosities (Sect. \ref{uvf}) are calculated at a
standard rest-frame wavelength for each source, in this case
\mbox{$\lambda_U=1216$}~\AA. As before (Sect. \ref{21-cm fluxes}), the
luminosity is given by the expression \mbox{$L_{\lambda_U} =
  4\pi\,D_{\rm QSO}^2 \, F_{\lambda_U(1+z)}/(z_{\rm em}+1)$}, although
rather than using a K-correction term, to correct a given
observed passband to the appropriate rest-frame wavelength, we instead
interpolate between or extrapolate from the observed photometry to
obtain the flux at the observed wavelength $\lambda_U(z+1)$.

The extrapolation or interpolation is done by fitting a power law
(i.e. a linear fit in $\log\lambda-\log F$ space). Thus, for two
observations at bands 1 and 2, with fluxes $F_1$ and $F_2$, the UV
flux is
\[
F_{\lambda_U(1+z)} = F_1 \left(\frac{\lambda_U(1+z)}{\lambda_1}\right)^\alpha,
{\rm~where~}
\alpha = \frac{\log(F_2/F_1)}{\log({\lambda_2/\lambda_1})}.
\]

Although the observational data we have obtained from the literature
is a heterogeneous mix of photometry, we try as much as possible to
maintain consistency by using the same bands. The combination that
covers the largest number of sources is $B$ and $R$ bands. Due to the
relatively low redshift of most of the sources, extrapolation from $B$
band is required. If $B$ is not available, $V$ is used instead, and if
one of $V$ or $R$ is not available, $K$ band was used. In one case,
J0924$-$2201, only $R$ was used since the high redshift meant that
$\lambda_U(1+z)$ fell into this band.

\label{lastpage}

\end{document}